\documentclass[12pt]{article}
\usepackage{pdfpages}
\usepackage{lmodern}
\usepackage{graphicx} % Required for including pictures 
\usepackage{wrapfig} % Allows in-line images
\usepackage{textcomp} % Allows in-line images
\usepackage[utf8]{inputenc}
\DeclareSymbolFont{letters}{OML}{ztmcm}{m}{it}
\usepackage{color}
\usepackage{bm}
\usepackage{multirow}

\usepackage{array}
\usepackage{listings}
\usepackage{adjustbox}% http://ctan.org/pkg/adjustbox
\usepackage[boxed]{algorithm2e}
\usepackage{subcaption}
\usepackage{multirow} 
\usepackage[OMLmathrm,OMLmathbf]{isomath}
\usepackage{booktabs} 
\usepackage{pdflscape}
\usepackage{natbib}
\usepackage{multibib}
\usepackage{multicol}
\newcites{supp}{Supplementary Material References}
\usepackage[T1]{fontenc} % Required for accented characters
\usepackage{tablefootnote}
\usepackage{amsmath}
\usepackage{etoolbox} 
\usepackage{amsthm}
\usepackage{amsfonts}
\usepackage{amssymb}
\usepackage[section]{placeins}
\usepackage{diagbox} 
\usepackage{hyperref}
\usepackage{graphicx} 
\usepackage{fixmath} 
\usepackage[scaled=.90]{helvet}
\usepackage{wrapfig}
\usepackage{tikz}
\usepackage{footnote}
\makesavenoteenv{tabular}
\makesavenoteenv{table}
\usepackage{setspace}
\usepackage[boxed]{algorithm2e}
\usepackage[symbol]{footmisc}

\usepackage{bbm}
\usepackage{bm}
\usepackage{caption}
\usepackage{subcaption}
\newcommand{\blind}{1}

\begin{document}

\def\spacingset#1{\renewcommand{\baselinestretch}%
{#1}\small\normalsize} \spacingset{1}

%%%%%%%%%%%%%%%%%%%%%%%%%%%%%%%%%%%%%%%%%%%%%%%%%%%%%%%%%%%%%%%%%%%%%%%%%%%%%%

\if1\blind
{
  \title{\bf Exponential Random Graph Models for Dynamic Signed Networks: An Application to International Relations}
  \author{Cornelius Fritz\thanks{
    The authors gratefully acknowledge support from the German Federal Ministry of Education and Research (BMBF) under Grant No. 01IS18036A and the German Research Foundation (DFG) for the project TH 697/11-1: Arms Races in the Interwar Period 1919-1939. Global Structures of Weapons Transfers and Destabilization. }\hspace{.2cm}\\
    Department of Statistics, LMU Munich\\
    Marius Mehrl \\
    Paul W. Thurner \\
    Geschwister Scholl Institute of Political Science \\
    Göran Kauermann\\
    Department of Statistics, LMU Munich
    }
  \maketitle
} \fi

\if0\blind
{
  \bigskip
  \bigskip
  \bigskip
  \begin{center}
    {\LARGE\bf  Exponential Random Graph Models for Dynamic Signed Networks: An Application to International Relations}
\end{center}
  \medskip
} \fi

\begin{abstract}

Substantive research in the Social Sciences regularly investigates signed networks, where edges between actors are either positive or negative. For instance, schoolchildren can be friends or rivals, just as countries can cooperate or fight each other. This research often builds on structural balance theory, one of the earliest and most prominent network theories, making signed networks one of the most frequently studied matters in social network analysis. While the theorization and description of signed networks have thus made significant progress, the inferential study of tie formation within them remains limited in the absence of appropriate statistical models. In this paper we fill this gap by proposing the Signed Exponential Random Graph Model (SERGM), extending the well-known Exponential Random Graph Model (ERGM) to networks where ties are not binary but negative or positive if a tie exists. Since most networks are dynamically evolving systems, we specify the model for both cross-sectional and dynamic networks. Based on structural hypotheses derived from structural balance theory, we formulate interpretable signed network statistics, capturing dynamics such as ``the enemy of my enemy is my friend''. In our empirical application, we use the SERGM to analyze cooperation and conflict between countries within the international state system. 

\end{abstract}

\noindent%


\vfill

\newpage

%	\noindent Highlights: 
%\begin{itemize}
  %  \item Introduction of Signed Exponential Random Graph models (SERGM)
  %  \item Extend state-of-the-art ERGM estimation and model selection methods to the SERGM 
 %   \item Implementation of the methods in the R package $\mathtt{ergm.sign}$
  %  \item Application to conflicts and alliances between societal groups and between states
%\end{itemize}

%\spacingset{1.9} % DON'T change the spacing!

\section{Introduction}
\label{sec:intro}
% Belarus, a strong ally of Russia, aided the invasion and continues preparations for an attack on Ukraine. 
%, but was also moved to close the Bosporus to warships of both sides, affecting mainly Russia, and sells armed drones to Ukraine
In February 2022, Russia invaded Ukraine. This invasion shifted the relations that numerous European countries had with the belligerents. The EU member states, including previously Russia-aligned countries such as Hungary, sanctioned Russia and provide support to Ukraine. Belarus, a close ally of Russia, followed its partner into the conflict and was accordingly also sanctioned by the EU member states. And Turkey, a political and economic partner of both Ukraine and Russia, struggled to remain neutral in the conflict and thus sought to mediate between the belligerents. A meaningful geopolitical adjustment thus followed the Russian attack which demonstrates the importance of positive \emph{and} negative ties in the international network of states, showing how pairwise cooperation and conflict between countries are interdependent.

Political scientists have studied this interplay of positive and negative ties between states since the early 1960s \citep{Harary_1961}. In this context, international relations are conceived as signed networks, where the nodes are states and the edges are either positive, corresponding to bilateral cooperation, negative, expressing bilateral conflict, or non-existent. Most of this research builds on structural balance theory, which postulates that triads are balanced if they include an odd number of positive relations and unbalanced if that number is either even (``strong'' structural balance; \citealp{Heider_1946,Cartwright_Harary_1956}) or exactly two (``weak'' structural balance; \citealp{Davis_1967}). Accordingly, International Relations scholars have studied whether specific triangular constellations correspond with these propositions \citep{Harary_1961,Healy_Stein_1973,McDonald_Rosecrance_1985,Doreian_Mrvar_2015} and what implications structural balance has for community formation and system polarization \citep{Hart_1974,Lee_Muncaster_Zinnes_1994}. More recently, studies %investigate the development of structural balance in the global interstate system over time \citep{Askarisichani_2020,Belaza_2017} and
seek to test whether structural balance affects interstate conflict and cooperation in an inferential framework \citep{Maoz_2007,Lerner_2016,Kinne_Maoz_2022}. %In this context, international relations are conceived as signed networks, where the nodes of the network are states and the edges are either positive, corresponding to bilateral cooperation, or negative, expressing a bilateral conflict or non-existent otherwise. 

% \citep{Lai2001}. Others also seek to explore how specific forms of signed relations affect conflict between states without explicitly adhering to structural balance theory \citep{Crescenzi_2007,Gartzke_Gleditsch_2021}.

However, the study of signed networks is not restricted to International Relations. There are also applications to friendship and bullying between children \citep{Huitsing_2012,Huitsing_2014}, alliances and conflicts between tribal \citep{Hage_Harary_1984} or criminal groups \citep{Nakamura_Tita_Krackhardt_2020}, statements of support and opposition between politicians \citep{Arinik_2020,DeNooy_Kleinnijenhuis_2013}, and even to interactions within ecological networks \citep{Saiz_2017}. In the setting of online social media and multiplayer games signed networks are also frequently studied \citep{Leskovec_2010,Bramson_Hoefman_Schoors_Ryckebusch_2021}. Signed networks are thus a substantively important subject of study across and beyond the Social Sciences. 

%There is also ongoing research on global (im-)balance measures for signed networks (e.g., \citealp{Facchetti_2011,Kirkley_2019}). 
When working with signed networks, most techniques known from binary networks are not directly appropriate. A significant amount of work thus focuses on adapting blockmodels \citep{Doreian2009,Jiang2015} as well as network statistics, such as  centrality  \citep{Everett2014} and status \citep{Bonacich2004}, to signed networks. From an inferential perspective, the study of signed networks so far has mainly relied on logistic regression \citep{Maoz_2007, Lerner_2016} or perceiving the observations as multivariate networks with multiple layers \citep{Huitsing_2012, Huitsing_2014,Stadtfeld2020}, where one level relates to the positive and another to the negative edges. While the former approach disregards endogenous dependence, the latter only allows for dependence between the separate observed layers of the network. Moreover, the multilayer approach does not adequately capture that most interactions in signed networks are either positive, negative, or non-existent. In other words, countries having negative and positive relations at the same time is unrealistic. % As a result, the inferential methods currently available to study signed networks lag behind what is available for binary networks. 
%there is a lack of tailored inferential frameworks for signed networks akin to what is available for binary networks. 

In the context of binary networks, \citet{frank1986} proposed Exponential Random Graph Models (ERGMs) as a generative model for a network encompassing $n$ actors represented by the adjacency matrix ${\mathbf{y}} = ({\mathbf{y}}_{ij})_{i,j = 1, ..., n}$, where $y_{ij} = 1$ translates to an edge between actors $i$ and $j$ and ${y}_{ij} = 0$ indicates that there is no edge. Henceforth, we use lowercase letters for variables when referring to the realized value of a random variable, i.e., the observed network $\mathbf{y}$, and capitalize the name to indicate that they are stochastic random variables, for instance, $\mathbf{Y}$. Within this framework, \citet{wasserman1996} formulate a probability distribution over all possible $\mathbf{y} \in {\mathcal{Y}}$ by a canonical exponential family model:
\begin{align}\label{eq:ergm}
\mathbb{P}_{\boldsymbol{\theta}}(\mathbf{Y} = \mathbf{y}) = \frac{\exp \left\{ \boldsymbol{\theta}^\top\mathbf{s}( \mathbf{y}) \right\}}{\kappa(\boldsymbol{\theta})} ~ \forall ~ \mathbf{y}  \in \mathcal{{Y}},
\end{align}
where $\mathcal{{Y}}$ is the set of all observable binary adjacency matrices among $n$ fixed actors, $s\text{ : }\mathcal{{Y}} \rightarrow \mathbb{R}^q$ is a function of sufficient statistics weighted by the coefficients $\theta \in \Theta\subseteq\mathbb{R}^p$, and  $\kappa(\boldsymbol{\theta}) := \sum_{\tilde{\mathbf{y}} \in \mathcal{{Y}}} \exp\{\theta^\top \mathbf{s}( \tilde{\mathbf{y}})\}$ is a normalizing constant.  Possible choices for the sufficient statistics $\mathbf{s}( \mathbf{{y}})$ of directed networks include the number of edges and triangles in the network (see \citealp{Lusher2012} for a detailed overview of the model and other possible statistics). Depending on the specific sufficient statistics, ERGMs relax the often unrealistic conditional independence assumption inherent to most standard regression tools in dyadic contexts and allow general dependencies between the observed relations. Note that in many applications, auxiliary information $\mathbf{x}$ exogenous to the network is available, which can also be used in the sufficient statistics. For brevity of the notation, we, however, omit the dependence of $\mathbf{s}$ on $\mathbf{x}$. Due to this ability to flexibly specify dependence among relations, account for exogenous information, the desirable properties of exponential families, and versatile implementation in the $\mathtt{ergm}$ $\mathtt{R}$ package \citep{Handcock2008,Hunter2008}, the ERGM is a core inferential approach in the statistical analysis of networks. %This framework has been extended to cover real-, count-, and rank-valued networks \citep{desmarais2012,krivitsky2012,Krivitsky2017}. 

In this article, we extend \eqref{eq:ergm} to cover signed networks under general dependency assumptions and coin the term Signed Exponential Random Graph Model (SERGM) for the resulting model. The SERGM provides an inferential framework to test the predictions of, e.g., structural balance theory \citep{Heider_1946,Cartwright_Harary_1956} without assuming that all observed relations are independent of one another. This characteristic is of vital importance given that balance theory explicitly posits that the sign of one relation depends on the state of other relations in the network. As the introductory examples suggest, interdependence-driven sign changes occur rapidly between states, necessitating the use of endogenous network statistics to adequately capture them. Along these lines, \citet[p.~75]{Lerner_2016} notes that ``tests of structural balance theory'' should not rely on ``models that assume independence of dyadic observations'' and thereby flags the importance of developing an ERGM for signed networks. We answer this call by introducing, applying, and, via the $\mathtt{R}$ package $\mathtt{ergm.sign}$, providing statistical software in $\mathtt{R}$ \citep{R} to implement the SERGM for static and dynamic networks, which is currently available at:
\begin{center}
\url{https://github.com/corneliusfritz/ergm.sign}    
\end{center}

%WHY ARE WE BETTER THAN ALL OTHER PEOPLE? 

%\begin{itemize}
  %  \item Networks are everywhere ... 
  %  \item Still most models can only deal with restrictive types of networks, i.e., binary ect. 
  %  \item But in this respect theory and data availability is ahead of the modeling possibilities -> More complex networks are observed but can hence not be studied in a principled inferential framework 
   % \item We come to signed networks -> why are they so important from a substantive point of view? 
    %\begin{itemize}
    %    \item From the PS angle 
    %    \item From the sociology and anthropology angle 
     %   \item What theories are here interesting? Structural Balance vs. Status Theory ect.
    %\end{itemize}
    %\item What has already been done for signed networks? 
    %\begin{itemize}
       % \item Look at the analyses from the SN SI (Stadtfeld, Lerner, ...)
       % \item Clustering was proposed and there are numerous articles in CS
   % \end{itemize}
   % \item What are we planning to do in this article? 
   % \begin{itemize}
    %    \item Use ERGMs 
%\item Introduce real quick ERGMs for binary data and motivate their use 
    %    \item What needs to be changed when the networks are signed? 
 %   \end{itemize}
%\end{itemize}

We proceed as follows: In the consecutive section, we formally introduce the SERGM and a novel suite of sufficient statistics to capture network topologies specific to signed networks. In Section  \ref{sec:estimation}, we detail how to estimate the parameters of the SERGM and quantify the uncertainty of the estimates. %Section \ref{sec:mod_sel} discusses how to evaluate the Akaike Information Criterion (AIC,\citealp{Akaike1974}) and adopt the model assessment procedures of \citet{Hunter2008b} to signed networks as two possibilities to carry out model selection. 
Next, we apply the introduced model class to the interstate network of cooperation and conflict in Section \ref{sec:application}. Finally, we conclude with a discussion of possible future extensions.

\section{The Signed Exponential Random Graph Model}
\label{sec:sergm}
\subsection{Model Formulation}
\label{sec:model_formulation}

%Since we will propose a probabilistic model, we refer to observations of random variables by lower case letters, e.g., $\mathbf{y}$ for the observed networks, and use upper case letters for the random variable, e.g., $\mathbf{Y}$ for the random variable describing the network.
%, although one might argue that self-loops are meaningless for signed networks in most real-world applications
%In analogy to the binary ERGM, we again omit the dependence on auxiliary information $\mathbf{x}$ in \eqref{eq:sergm}. 
First, we establish some notation to characterize signed networks. Assume that the signed adjacency matrix $\mathbf{y} = (y_{ij})_{i,j = 1, ..., n}$ was observed between $n$ actors. Contrasting the binary networks considered in \eqref{eq:ergm}, the entries of this signed adjacency matrix  $y_{ij}$ are either ``$+$'', ``$-$'', or ``0'', indicating a positive, negative, or no edge between actors $i$ and $j$. To ease notation, we limit ourselves to undirected networks without any self-loops, i.e., $\forall~ i,j = 1, ..., n ~ y_ {ij} = y_ {ji}$ and $y_{ii} =$ ``0'' holds. Nevertheless, the proposed model naturally extends to directed settings. We denote the space encompassing all observable signed networks between $n$ actors by $\mathcal{Y}^\pm$ and specify a distribution over this space analogous to \eqref{eq:ergm} in the following log-linear form:
\begin{align}\label{eq:sergm}
\mathbb{P}_{\boldsymbol{\theta}}(\mathbf{{Y}} =\mathbf{{y}}) = \frac{\exp \left\{ \boldsymbol{\theta}^\top s( \mathbf{{y}}) \right\}}{\kappa(\boldsymbol{\theta})}  ~ \forall ~ \mathbf{y} \in \mathcal{Y}^\pm.
\end{align}
The function of sufficient statistics in \eqref{eq:sergm} takes a signed network as its argument and determines the type of dependence between dyads in the network. A theoretically motivated suite of statistics one can incorporate as sufficient statistics follows in Section \ref{sec:suff_statistics} but mirroring the term counting edges in binary networks, we can use the count of positive ties in signed network $\mathbf{y}$ via 
\begin{align*}
    EDGE^+( \mathbf{{y}}) = \sum_{i<j} \mathbb{I}(y_{ij} = \text{``+''}),
\end{align*}
where $\mathbb{I}(\cdot)$ is the indicator function. Along the same lines, one can define a statistic for the number of negative edges $EDGE^-( \mathbf{{y}})$ and use both statistics as intercepts in the model. %Additionally, the size of the unconstrained sample space of undirected signed network without loops is $\left(\frac{n(n-1)}{2}\right)^3$ as opposed to $\left(\frac{n(n-1)}{2}\right)^2$ for the binary case. 

We can extend \eqref{eq:sergm} to dynamic networks, which we denote by $\mathbf{Y}_1, ..., \mathbf{Y}_T$ for observations at $t = 1, ..., T$, by assuming a first-order Markov dependence structure to obtain
\begin{align}\label{eq:tsergm}
\mathbb{P}_{\boldsymbol{\theta}}(\mathbf{Y}_t = \mathbf{y}_t|\mathbf{Y}_{t-1} =  \mathbf{y}_{t-1}) = \frac{\exp \left\{ \boldsymbol{\theta}^\top\mathbf{s}( \mathbf{y}_{t}, \mathbf{y}_{t-1}) \right\}}{\kappa(\boldsymbol{\theta}, \mathbf{y}_{t-1})} ~ \forall ~ \mathbf{y}_t \in \mathcal{Y}^\pm.
\end{align} 
The sufficient statistics encompassed in $\mathbf{s}(\mathbf{y}_{t}, \mathbf{y}_{t-1})$ capture within-network or endogenous dependencies through statistics that only depend on $\mathbf{y}_{t}$ and between-network dependencies when incorporating $\mathbf{y}_{t-1}$. One instance for network statistics for between-network dependency is the stability statistic for positive edges  
\begin{align*}
    STABILITY^+(\mathbf{{y}}_t,\mathbf{{y}}_{t-1}) = \sum_{i<j} \mathbb{I}(y_{ij,t} = \text{``+''})\mathbb{I}(y_{ij,t-1} = \text{``+''}),
\end{align*}
which can equivalently be defined for negative ties. Thus, we assume that the observed network is the outcome of a Markov chain with state space $\mathcal{Y}^\pm$ and transition probability \eqref{eq:tsergm}. Of course, we may also include exogenous terms in \eqref{eq:tsergm}, i.e., any pairwise- or actor-specific information external to $\mathbf{y_t}$.%, and  endogenous terms, which depend on structural characteristics of $\mathbf{y}_t$.

%Due to this temporal setting, we can now also incorporate any statistics making use . 

%Although \eqref{eq:tsergm} might at first seem equivalent to \eqref{eq:ergm}, we stress that the support of the distribution is the space of all signed networks. 

For the interpretation of the estimates, techniques from binary ERGMs can be adapted. To derive a local tie-level interpretation, let $\theta_q$ with $q \in \{1, ..., p\}$ denote the $q$th entry of $\theta$ corresponding to the $q$th sufficient statistic, $s_{q}( \mathbf{y}_t, \mathbf{y}_{t-1})$. We further define ${\mathbf{y}_t} = (y_{ij,t})_{i,j = 1, ..., n}$ for $t = 1, ..., T$ and by $\mathbf{y}_{ij,t}^+$ denote the network $\mathbf{y}_t$ with the entry $y_{ij,t}$ fixed at ``$+$'', $\mathbf{y}_{ij,t}^-$ and $\mathbf{y}_{ij,t}^0$ are established accordingly. Let $\mathbf{y}_{(-ij),t}$ refer to the network $\mathbf{y}_t$ excluding the entry $y_{ij,t}$. Due to the added complexity of signed networks, the distribution of $Y_{ij,t}$ conditional on $\mathbf{Y}_{(-ij),t}$ is a multinomial distribution where the event probability of entry ``+'' is: 
\begin{equation}
\begin{aligned}
    \mathbb{P}_\theta&(Y_{ij,t} =\text{``+''} | \mathbf{Y}_{(-ij),t} = \mathbf{y}_{(-ij),t-1}, \mathbf{Y}_{t-1} = \mathbf{y}_{t-1})  = \frac{\exp \left\{ \boldsymbol{\theta}^\top\mathbf{s}( \mathbf{y}_{ij,t}^+,\mathbf{y}_{t-1}) \right\}}{\sum_{k \in \{+,-,0\}}\exp \left\{ \boldsymbol{\theta}^\top\mathbf{s}( \mathbf{y}_{ij,t}^k,\mathbf{y}_{t-1}) \right\} }.    \label{eq:conditional_prob}
    \end{aligned}
\end{equation}
In the same manner, we can state the conditional probability of ``$-$'' and ``0''. In accordance with {change statistics} from binary ERGMs, we subsequently define {positive} and {negative change statistics} through 
\begin{equation}
\begin{aligned}
     \label{eq:change_stats}
   \Delta_{ij,t}^{0\rightarrow +}(\mathbf{y}_{(-ij),t}, \mathbf{y}_{t-1}) &= \mathbf{s}( \mathbf{y}_{ij,t}^+, \mathbf{y}_{t-1}) - \mathbf{s}( \mathbf{y}_{ij,t}^0, \mathbf{y}_{t-1}) \\
  \Delta_{ij,t}^{0\rightarrow -} (\mathbf{y}_{(-ij),t}, \mathbf{y}_{t-1}) &= \mathbf{s}( \mathbf{y}_{ij,t}^-, \mathbf{y}_{t-1}) - \mathbf{s}( \mathbf{y}_{ij,t}^0, \mathbf{y}_{t-1}).
\end{aligned}
\end{equation}
While the  {positive change statistic} $\Delta_{ij,t}^{0\rightarrow +}( \mathbf{{y}}_t, \mathbf{y}_{t-1})$ is the change in the sufficient statistics resulting from flipping the edge value of $y_{ij,t}$ from ``0'' to ``+'',  the {negative change statistic} $\Delta_{ij,t}^{0\rightarrow -} (\mathbf{{y}}_t, \mathbf{y}_{t-1})$ relates to the change from ``0'' to ``$-$''. By combining \eqref{eq:conditional_prob} and \eqref{eq:change_stats}, we can obtain the relative log odds of $Y_{ij,t}$ to be ``+'' and ``$-$'' rather than ``0'': 
\begin{equation}
 \label{eq:relative_risks}
\begin{aligned}
    \log\left(\frac{\mathbb{P}_\theta(Y_{ij,t} = \text{``+''} | \mathbf{Y}_{(-ij),t} = \mathbf{y}_{(-ij),t}, \mathbf{Y}_{t-1} =\mathbf{y}_{t-1})}{\mathbb{P}_\theta(Y_{ij,t} =\text{``0''}  | \mathbf{Y}_{(-ij),t} = \mathbf{y}_{(-ij),t}, \mathbf{Y}_{t-1} =\mathbf{y}_{t-1})} \right) &=\theta^\top\Delta_{ij,t}^{0\rightarrow +}( \mathbf{y}_{(-ij),t}, \mathbf{y}_{t-1}) \\
      \log\left(\frac{\mathbb{P}_\theta(Y_{ij,t} = \text{``$-$''} | \mathbf{Y}_{(-ij),t} = \mathbf{y}_{(-ij),t} , \mathbf{Y}_{t-1} = \mathbf{y}_{t-1})}{\mathbb{P}_\theta(Y_{ij,t} = \text{``0''}  | \mathbf{Y}_{(-ij),t} = \mathbf{y}_{(-ij),t}, \mathbf{Y}_{t-1} = \mathbf{y}_{t-1})} \right) &= \theta^\top\Delta_{ij,t}^{0\rightarrow -}( \mathbf{y}_{(-ij),t}, \mathbf{y}_{t-1}).
\end{aligned}
\end{equation}
This allows us to relate $\theta$ to the conditional distribution of $Y_{ij,t}$ given the rest of the network and derive two possible interpretations of the coefficients reminiscent of multinomial and logistic regression: the conditional log-odds of $Y_{ik,t}$ to be ``$+$'' rather than ``$0$'' are changed by the additive factor $\theta_p$, if the value of $y_{ij,t}$ changing  from ``0'' to ``$+$'' raises the $p$th entry of $\Delta_{ij,t}^{0\rightarrow +}( \mathbf{y}_{(-ij),t}, \mathbf{y}_{t-1})$ by one unit, while the other statistics remain unchanged. A similar interpretation holds for the {negative change statistic}. 
%Depending on the statistic one wants to interpret either the interpretation  relying on the {positive} or {negative change statistic} is more insightful. For instance, $s_{\mathtt{edges\_pos}}( \mathbf{{y}})$ only affects $ \Delta_{ij}^{0\rightarrow 1}( \mathbf{{y}})$, which is hence more interesting to analyze, while the contrary holds for $s_{\mathtt{edges\_neg}}( \mathbf{{y}})$. 

Second, one can employ a global interpretation to understand the parameters on a network level. Then, $\theta_q>0$ indicates that higher values of $s_{q}( \mathbf{{y}}_t, \mathbf{y}_{t-1})$ are expected under \eqref{eq:sergm} than under a multinomial graph model, which we define as a simplistic network model where the value of each dyad is ``+'', ``$-$'' and ``0'' with equal probability. In the opposing regime with $\theta_q<0$, we expect lower values than under this multinomial graph model.

\subsection{From Structural Balance Theory to Sufficient Statistics}
\label{sec:suff_statistics}

As discussed in the introduction, structural balance theory is a natural approach to signed networks. But so far, inferential work on it remains limited and uses, as we show below, suboptimal measures of its structural expectations. We thus shortly introduce the core logic of structural balance theory, discuss previous measures of it, and then derive sufficient statistics from it for inclusion in the SERGM. These statistics enable us to test the structural expectations formulated by structural balance theory in a principled manner within the framework introduced in Section \ref{sec:model_formulation}.
 
%Against the background of increased interest in structural balance theory for the study of signed networks, it is of vital interest to represent the implications of this theory with the model framework introduced in Section \ref{sec:model_formulation}. Therefore, we sketch the network theory to    

\paragraph*{Theory}
%\textcolor{red}{STILL UNDER CONSTRUCTION AND TO BE FINISHED}
%\begin{itemize}
    %\item Start with how the choice of sufficient statistics implies specific types of dependency structures on the dyadic level 
    %\item What statistics can be of interest when working with signed networks? 
    %\item Start with the classics: 
    %\begin{enumerate}
    %    \item Positive and negative degree statistics 
    %    \item Balanced and imbalanced triangles 
    %\end{enumerate}
    %\item Problem of degeneracy when working with the triangle type of statistics -> cite the classics from ERGM 
    %\item To circumvent this behavior \citet{snijders2006} proposed the geometrically weighted statitstics -> we next extend those statistics to SERGM 
    %\item In the main article the focus should be on the edgewise statistics -> dyadwise statistics maybe for the Annex? Would otherwise maybe be a bit too much
   % \item Introduce GWDEG, GWESF and GWESE 
    %\item Section should probably be 1-2 pages? 
%\end{itemize}

The main implication of structural balance theory relates to the existence of triads between actors. Triads are the relations between three actors \citep{Wasserman1994} and generally called balanced if they consist solely of positive ties (``the friend of my friend is my friend'') or one positive and two negative ties (``the enemy of my enemy is my friend''). According to structural balance theory, this type of triad should be observed more often than expected by chance in empirical signed networks. In contrast, triads that include a single negative tie are structurally imbalanced as the node participating in both positive relations has to cope with the friction of its two ``friends'' being opposed to each other. This actor should thus try to turn the negative tie into a positive tie to achieve a balanced constellation where all three actors share positive connections. But if this proves impossible, the actor will eventually have to choose a side, making one of its previously positive ties negative and resulting in structural balance. 
%In triads where a single negative relation exists, there is thus a clear pressure on one of the actors to either manufacture a change of this relation or, instead, shift one of its own ties from positive to negative. 
In triads where relations between all three actors are negative, the actors at least have incentives to make similar changes; these triads are thus also considered structurally imbalanced \citep{Heider_1946,Cartwright_Harary_1956}. In particular, two actors could reap benefits by developing a positive relationship, pooling their resources, and ganging up on the third node. However, later work views these triads without any positive ties as weakly balanced \citep{heider1958psychology,Davis_1967}, as \citet{Davis_1967} notes that enemies of enemies being enemies indicates structural imbalance only if there are two subsets of nodes in the network. Triadic constellations with one negative relation are thus structurally imbalanced, should be empirically rare, and, where they exist, tend to turn into balanced states. Where only one negative tie exists, there is strong pressure to either eliminate it or create an additional one. And where there are three negative ties, actors at least have a clear incentive to turn one of them into a positive relation opportunistically, though their (im-)balance depends on the nature of the wider system (see also \citealp[ch.~5]{Easley_Kleinberg_2010}).

\paragraph*{Testing Structural Balance via Lagged Statistics}

\begin{figure}[t!]
     \centering
     \includegraphics[width=0.73\textwidth, page = 2,trim={0cm 0cm 0cm 0cm},clip]{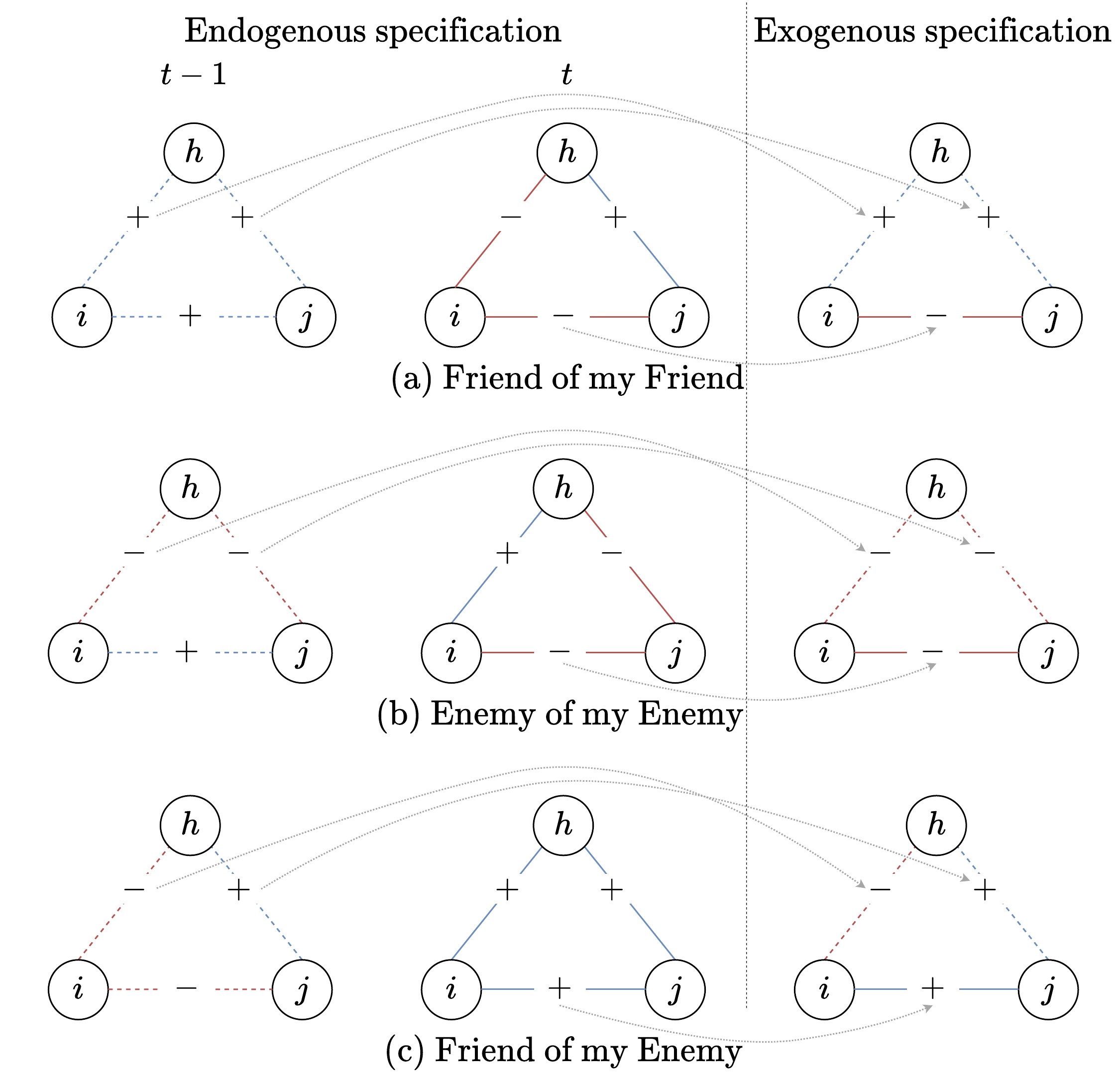}
        \caption{Combining past and present ties can misrepresent structural (im-)balance: Triads observed at $t-1$ and $t$ are balanced (left side), combined triads are imbalanced (right side). Dashed lines indicate tie at $t-1$, solid ones at $t$. Dotted arrows show which ties from $t-1$ and $t$ contribute to the exogenous specification.}
        \label{fig:exogversions}
\end{figure}

In interstate relations, this theory implies that two countries that are on friendly terms with the same other state should not wage war against each other. If three states all engage in conflict with each other, two of them may also find it beneficial to bury their hatchet, focus on their common enemy, and pool their resources against it. Along these lines, existing research asks whether two countries' probability to cooperate or to fight is affected by them sharing common friends or foes \citep{Maoz_2007, Lerner_2016}. In particular, these authors investigate whether having shared allies or enemies at time $t-1$ affects the presence of positive and negative ties at $t$. The resulting ``friend of my friend is my friend'' statistic we can incorporate in the sufficient statistics of \eqref{eq:tsergm} is:
\begin{align}
    \label{eq:stat_fof}
    CF^+(\mathbf{y}_t, \mathbf{y}_{t-1}) = \sum_{i<j} \mathbb{I}(y_{ij,t} =\text{``+''}) \left( \sum_{\substack{h \neq i, h \neq j}} \mathbb{I}(y_{ih,t-1} = \text{``+''}) \mathbb{I}(y_{jh,t-1} =\text{``+''})\right).
    \end{align}
Similar delayed statistics can be defined for all other implications of the theory by treating the existence of common friends and foes as exogenous covariates. However, this approach comes with both theoretical and methodological problems. It is unclear whether actors wait a period (a calendar year in the case of \citealp{Maoz_2007} and \citealp{Lerner_2016}) to adjust their relations towards structural balance and why they should do so as other applications of structural balance theory view these changes as instantaneous \citep[see e.g.][]{Kinne_Maoz_2022}. If the countries do not wait for a period, this approach can misrepresent the dynamics of signed networks as contradicting structural balance theory when they do not. 

To illustrate this point, the right side of Figure \ref{fig:exogversions} visualizes three structurally imbalanced constellations which \citet{Maoz_2007} and \citet{Lerner_2016} uncover in the network of cooperation and conflict between states: (a) The friend of a friend being an enemy, (b) the enemy of an enemy being an enemy, and (c) the friend of an enemy being a friend. The left side of Figure \ref{fig:exogversions} presents the triads at $t-1$ and $t$ these constellations are potentially made up of as ties are not observed simultaneously. The links of $i$ and $j$ to $h$ were observed at $t-1$ but those between $i$ and $j$ at $t$. The structurally imbalanced triads on the right side of Figure \ref{fig:exogversions} %, as documented by \citet{Maoz_2007} and \citet{Lerner_2016}, 
thus consist of observations of the same triad made at two different points in time. Crucially, the left side of Figure \ref{fig:exogversions} shows that both of these observations can themselves be structurally balanced. Exogenous measures of common friends and enemies can thus only capture the predictions of structural balance theory if (i) actors $i$ and $j$ wait a period until they change their tie sign due to their links to $h$ and (ii) their links to $h$ remain unchanged. Both of these conditions require strong assumptions regarding how actors behave within a network. In particular, structural balance theory implies that the edges between $i$, $j$, and $h$ are interdependent. But its exogenous operationalization assumes two of these edges as fixed while waiting to observe the third. An example shows that this is not just a theoretical issue, but mischaracterizes empirically observed relations between states: The US and Iran had common foes in 1978 but, in 1979, had become outright enemies themselves. The exogenous operationalization of structural balance regards this situation as unbalanced although it is an example of the scenario of Figure \ref{fig:exogversions}b. 
%In 1978, Iran was ruled by the Shah and a staunch ally of the US against Soviet-backed Arab countries. They were thus allies sharing common enemies. But the Shah was disposed in January 1979, eventually being replaced by a theocratic government that saw the US not as an ally but an enemy. In 1979, the US--Iran tie thus had become negative and while Iran remained opposed to many of its Arab regional rivals, the US actually struck up positive relations with some of them, e.g. Egypt due to the Camp David accords. As such, US--Iran relations were part of balanced structures both in 1978 and 1979 even though the exogenous measure of structural balance would classify them as imbalanced in 1979. More generally, ruling coalition changes, such as the Iranian Revolution of 1979, are associated with foreign policy shifts \citep{mattes2015leadership,Pilster_Bohmelt_Tago_2015}. US--Iran relations are thus a prominent but not the only example of mischaracterized imbalance as presented in figure \ref{fig:exogversions}.       

%The signed relation between $i$ and $j$ depends on their respective relations with $h$, but this is not a one-way street. The relation of $h$ with $j$ is similarly affected by the signs of the ties $ij$ and $ih$. As a result we see that fixing the ties between $i$ and $h$ and $j$ and $h$ while ``waiting'' for their effect on $ij$ is inappropriate. 

\paragraph*{Testing Structural Balance via Endogenous Statistics}
\begin{figure}[t!]
     \centering
     \includegraphics[width=0.3\textwidth, page = 1]{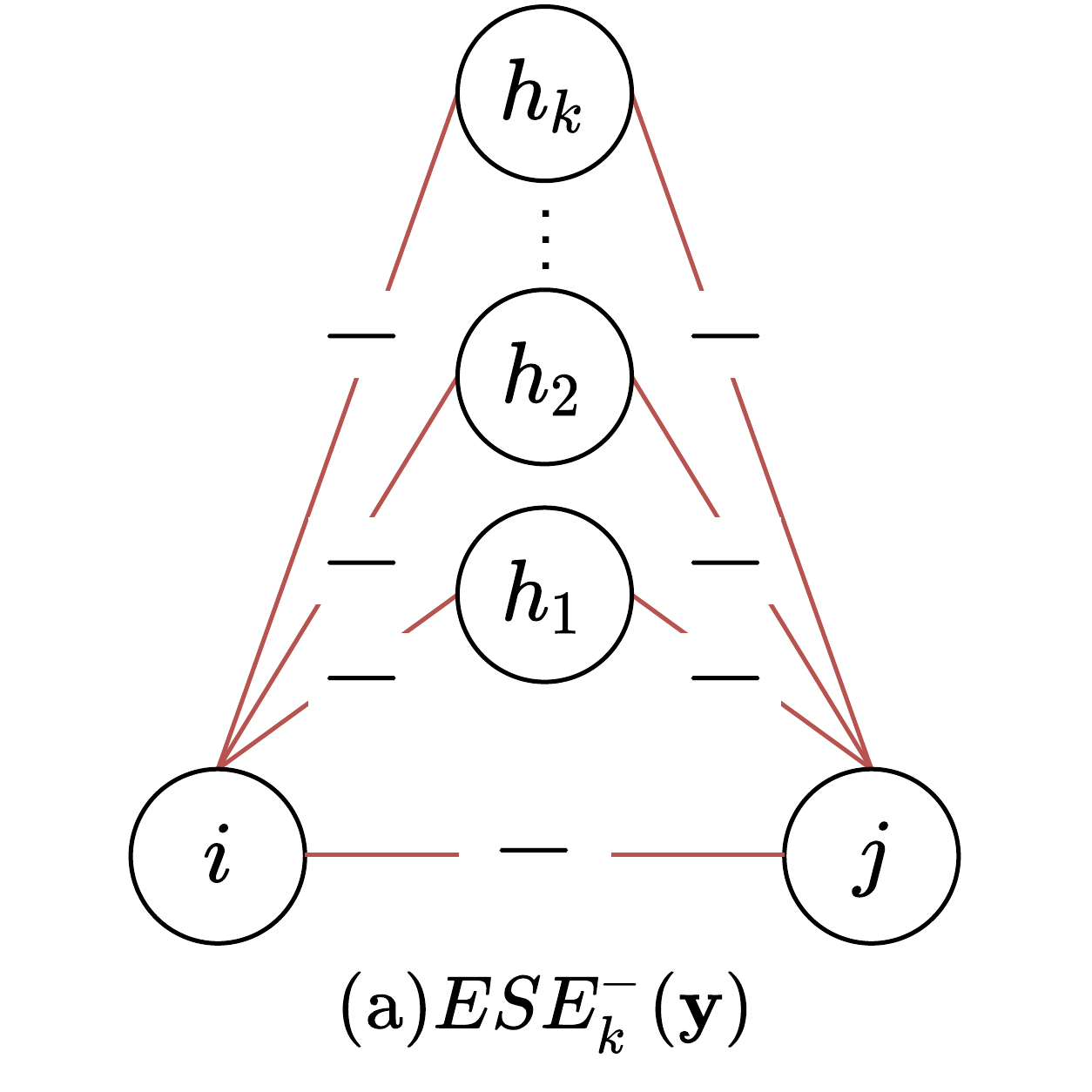}
     \includegraphics[width=0.3\textwidth, page = 2]{Plots/plot_alt_2.pdf} \\
     \includegraphics[width=0.3\textwidth, page = 4]{Plots/plot_alt_2.pdf}
     \includegraphics[width=0.3\textwidth, page = 5]{Plots/plot_alt_2.pdf}
        \caption{Sufficient statistics for signed networks.}
        \label{fig:suff_stats}
\end{figure}
%More fundamentally, treating triadic constellations as exogenous, and thus assuming ties between actors to be conditionally independent from each other, arguably goes against the core idea of structural balance theory that these edges are interdependent. The signed relation between $i$ and $j$ depends on their respective relations with $h$, but this is not a one-way street as $h$'s relation with $j$ is similarly affected by the signs of the ties $ij$ and $ih$. 
Therefore, endogenous network terms are necessary to capture the endogenous network dynamics postulated by structural balance theory. We next define endogenous statistics that mirror each constellation described by structural balance theory to test its predictions empirically.   
%From the perspective of substantive theory, the most crucial statistics for signed networks are triangles. Structural balance theory focuses on triangles between three actors, positing that ``the enemy of my enemy is my friend'' or ``the friend of my friend is my friend'', and we define signed triadic structures following these predictions. 
Building on the $k$-Edgewise-Shared Partner statistic developed to measure transitive closure in binary ERGMs \citep{hunter2007curved}, we can define $k$-Edgewise-Shared Friends, $ESF_k(\mathbf{y})$, and $k-$Edgewise-Shared Enemies, $ESE_k(\mathbf{y}_t)$, for signed networks. $ESF_k(\mathbf{y}_t)$ counts the edges with $k$ shared friends and $ESE_k(\mathbf{y}_t)$ those with $k$ shared enemies. We further differentiate between the state of the edge at the center of each triangular configuration and, e.g., write $ESF^+_k(\mathbf{y}_t)$ and $ESF^-_k(\mathbf{y}_t)$ as the version of the statistic where the value of $y_{ij}$ is ``+'' and ``$-$'', respectively. Figure \ref{fig:suff_stats} illustrates the resulting four statistics. %For instance, in Figure \ref{fig:suff_stats} (a) there is an negative edge between actors $i$ and $j$ and they share $k$ enemies; hence if Figure \ref{fig:suff_stats} (a) would be the complete network, $ESF^+_k(\mathbf{y})$ = 1 would hold. 

%, and we accordingly also develop geometrically weighted versions of their signed equivalents
For $k = 2$ these statistics reduce to specific types of common triangle configurations \citep{Holland1972}. However, as shown in \citet{snijders2006}, these types of statistics frequently lead to degenerate distributions where most of the probability mass is put on the empty or full graph \citep{handcock2003assessing,schweinberger2011instability}. Moreover, the implied {avalanche effect} is particularly pronounced if the corresponding parameters are positive, as structural balance theory suggests. For binary ERGMs, it is thus standard to employ a statistic of the weighted sum of statistics in which the weights are proportional to the geometric  sequence \citep{snijders2006,hunter2006}. We follow this practice and define the geometrically weighted statistic for negative edgewise-shared enemies, as portrayed in Figure \ref{fig:suff_stats}a, with a fixed decay parameter $\alpha$ as 
\begin{align}
    \label{eq:stat_gwese}
    GWESE^+(\mathbf{y}_t,\alpha) = \exp\{\alpha\} \sum_{k = 1}^{n-2} \left(1-\left(1-\exp\{-\alpha\}\right)\right)^k ESE^+_k(\mathbf{y}_t).
\end{align}
We establish the geometrically weighted variants of $ESE^-_k(\mathbf{y}_t),  ESF^+_k(\mathbf{y}_t),$ and $ESF^-_k(\mathbf{y}_t)$ accordingly. Each of these statistics reflects a specific type of triadic closure in signed networks as visualized in Figure \ref{fig:suff_stats}. To interpret the coefficient $\theta_{GWESE^+}$ one can consider the logarithmic relative change in the probability according to \eqref{eq:tsergm} when increasing the number of common enemies of a befriended edge by one and keeping all other statistics constant. If the befriended actors already had $k$ prior common enemies before this change, this relative change is given by
\begin{align*}
    \theta_{GWESE^+} \left(1-\left(1-\exp\{-\alpha\}\right)\right)^k.
\end{align*}
Thus, if $\theta_{GWESE^+}>0$, each additional common enemy raises the probability to observe the signed network, although the increments become smaller for higher values of $k$. \citet{hunter2007curved} shows that these geometrical weighted statistics are equivalent to the alternating $k$-triangle statistics proposed by \citet{snijders2006}.

%hypothesized mechanism that, due to structural balance effects, $i$ and $j$ are friends when they share common friends or common enemies. But, when focusing on a negative instead of a positive edge between {i} and {j}, they can also capture the idea that enemies of enemies will fight each other (see \citealp{Maoz_2007}). %As shown in figure \ref{fig:suff_stats}, ESE and ESF thus exist for both positive and negative focal ties. These sufficient statistics are substantively informed by structural balance theory while presenting natural extensions of statistics known from the statistical analysis of binary networks. 

%In the context of signed networks, ESE($\mathbf{y}$) and ESF($\mathbf{y}$) are endogenous statistics which can clearly be motivated by theoretical considerations. 

These triadic structures fully capture the logic of structural balance as they allow us to study the prevalence of triads where positive ties account for zero (Figure \ref{fig:suff_stats}a) , one (Figure \ref{fig:suff_stats}b), two (Figure \ref{fig:suff_stats}c), and all three (Figure \ref{fig:suff_stats}d) of the edges. According to this logic, we would expect the statistics $GWESE^+(\mathbf{y}_t$) and $GWESF^+(\mathbf{y}_t$) to be higher in empirical networks than expected by chance, but not $GWESE^-(\mathbf{y}_t$) and, particularly, $GWSF^-(\mathbf{y}_t$). If, on the other hand, the coefficients corresponding to $GWESE^-(\mathbf{y}_t$) or $GWESF^-(\mathbf{y}_t$) turn out to be positive in a network, this would offer empirical support for modifications of structural balance theory that also see the constellation in Figure \ref{fig:suff_stats}a as balanced \citep{heider1958psychology,Davis_1967} or combine it with insights about, e.g., opportunism or reputation \citep{Maoz_2007}. Mirroring the development of edge-wise shared enemy and friend statistics, it is also possible to compute dyad-wise statistics that do not require $i$ and $j$ to share a tie. 

%We focus on statistics for undirected signed networks here as they are appropriate to test structural balance theory. However, the endogenous terms defined here can be extended to the directed case, e.g., when seeking to test the predictions of status theory \citep{Leskovec_2010}. 

\paragraph*{Other Sufficient Statistics}

Besides these substantively informed statistics developed from structural balance theory, there are -  as in the binary case -  numerous other statistics one may incorporate into the model. Some of these are even necessary to isolate the effects of structural balance. In binary networks, closed triads where each node is connected to the others are more likely to form if the involved actors are highly active due to processes such as popularity. In the context of ERGMs, this phenomenon is captured by degree statistics counting the number of actors in the network with a specific number of edges. For signed networks, similar but more complicated processes may be at work and, to capture them, we define $DEG^+_k(\mathbf{y}_t)$ and $DEG^-_k(\mathbf{y}_t)$ as statistics that, respectively, count the number of actors in the signed network $\mathbf{y}_t$ with degree $k\in \{1, ..., n-1\}$ for ``+''- and ``$-$''-signed links, respectively. Since the degree statistics are also prone to the degeneracy issues detailed above, we define geometrically-weighted equivalents for the positive and negative degrees. One can also incorporate exogenous statistics for the propensity to observe either a positive tie, similar to \eqref{eq:stat_fof}, via the following statistic: 
\begin{align*}
    EXO^+(\mathbf{y}_t) &= \sum_{i<j} \mathbb{I}(y_{ij,t} = \text{``+''})x_{ij,t},
    \end{align*}
where $x_{ij,t}$ can be any pairwise scalar information. Similar statistics can be defined for negative, $EXO^-(\mathbf{y}_t)$, and any , $EXO^\pm(\mathbf{y}_t)$, tie. To test whether there is a tendency for homo- or heterophily based on actor attribute $x = (x_1, ..., x_n)$ in the network, one may transform the nodal information to the pairwise level by setting $x_{ij,t} = |x_{i,t} - x_{j,t}|$ or  $x_{ij,t} = \mathbb{I} (x_{i,t} = x_{j,t})$ for continuous and categorical attributes, respectively.

\section{Estimation and Inference}
\label{sec:estimation}

To estimate $\theta$ for a fully specified set of sufficient statistics, we maximize the likelihood of \eqref{eq:tsergm} conditional on the initial network $\mathbf{y}_0$:
\begin{align}
    \mathcal{L}(\theta; \mathbf{y}_1, ...,\mathbf{y}_T) =& \prod_{t = 1}^T  \frac{\exp \left\{ \boldsymbol{\theta}^\top\mathbf{s}( \mathbf{y}_{t}, \mathbf{y}_{t-1}) \right\}}{\kappa(\boldsymbol{\theta}, \mathbf{y}_{t-1})} 
    \label{eq:clh}
    =\frac{    \exp\left\{ \boldsymbol{\theta}^\top \left(\sum_{t = 1}^T \mathbf{s}( \mathbf{y}_{t}, \mathbf{y}_{t-1}) \right)\right\}}{\prod_{t = 1}^T \kappa(\theta,\mathbf{y}_{t-1})}.
\end{align}
%in what follows we shorten $\sum_{t = 1}^T \mathbf{s}( \mathbf{y}_{t}, \mathbf{y}_{t-1})$ to $\tilde{s}(\mathbf{y}_{1}, ..., \mathbf{y}_{T})$.
We can observe that this joint probability of the observed networks is still an exponential family, where the sufficient statistic is the sum of the individual statistics, the normalizing constant is composed of the product of the normalizing constants at each time point, and the canonical parameter is unchanged. Evaluating the normalizing constant in \eqref{eq:clh}, on the other hand, necessitates the calculation of  $T\cdot \left( 3^{\frac{n(n-1)}{2}}\right)$ summands, making the direct evaluation of the likelihood prohibitive even for small networks. Fortunately, these difficulties are known from the analysis of binary networks and have been tackled in numerous articles (see, e.g., \citealp{strauss1990pseudolikelihood,Hummel2012,Snijders2002,hunter2006}), which guide our estimation approach for the SERGM.

To circumvent the direct evaluation of \eqref{eq:clh}, we can write the logarithmic likelihood ratio of $\theta$ and a fixed $\theta_0$ without a normalizing constant but an expected value
\begin{equation}
\begin{aligned}
 r(\theta,\theta_0;\mathbf{y}) =  &(\theta - \theta_0)^\top \left(\sum_{t = 1}^T \mathbf{s}(\mathbf{y}_t, \mathbf{y}_{t-1})\right) \\&-\log\left(\mathbb{E}_{\theta_0}\left(\exp\left\{(\theta - \theta_0)^\top \left(\sum_{t = 1}^T \mathbf{s}(\mathbf{Y}_t, \mathbf{y}_{t-1})\right)\right\}\right)\right).
 \label{eq:lhr}
\end{aligned}
\end{equation}
%where $\mathbb{E}_\theta(f(\mathbf{Y}))$ is the expected value of the random variable $\mathbf{Y}$ under \eqref{eq:sergm} and $\theta$ transformed by the arbitrary function $f(\cdot)$ taking values from $\mathcal{Y}^\pm$ as its arguments. 

%\begin{algorithm}[t!]
%	\SetAlgoLined
%	\KwResult{$\mathbf{y}^{(1)}, ..., \mathbf{y}^{(K)}$}
%	\emph{Set}  $\mathbf{y}^{(0)},\theta$ and $K$\\
%	\emph{Set}  $\mathbf{y} = \mathbf{y}^{(0)}$\\
%	\For{$k = 1,..., K$}{
%		\emph{Select dyad} $(i^{(k)}, j^{(k)})$ randomly from all possible dyads\\
%		\emph{Set}  $\mathtt{cur} = y^{(k-1)}_{i^{(k)}, j^{(k)}}$\\
%		\emph{Sample}  $Y_{i^{(k)}, j^{(k)}} | \mathbf{Y}_{i^{(k)}, j^{(k)}}^C$ from $\{1,-1,0\}$ with event probabilities given in \eqref{eq:event_proba}\\
%		\emph{Set}  $\mathbf{y}^{(k)} =\mathbf{y}$\\
%	}
%	\caption{Gibbs sampler for signed networks under \eqref{eq:sergm}.}
%	\label{alg:gibbs}
%\end{algorithm}

We approximate the expectation in \eqref{eq:lhr} by sampling networks over time, denoted by $\mathbf{Y}^{(m)} = (\mathbf{Y}_1^{(m)}, ..., \mathbf{Y}_T^{(m)})$ for the $m$th sample, whose dynamics are governed by \eqref{eq:tsergm} under $\theta_0$. Due to the Markov assumption, it suffices to specify only how to sample $\mathbf{Y}_t^{(m)}$ conditional on $\mathbf{y}_{t-1}$ for $t = 1, ..., T$ via Gibbs sampling. In particular, we generate a Markov chain with state space $\mathcal{Y}^\pm$ that, after a sufficient burn-in period, converges to samples from $\mathbf{Y}_t$ conditional on $\mathbf{y}_{t-1}$. Since we toggle one dyad in each iteration, the conditional probability distribution we sample from is the multinomial distributions stated in \eqref{eq:conditional_prob}. In a setting where we sample $Y_{ij,t}$ conditional on $\mathbf{y}_{(-ij),t}$ and $ \mathbf{y}_{t-1}$ with its present value given by $\tilde{y}_{ij,t}$, we can restate this conditional probability for ``$+$'' in terms of change statistics:
\begin{align*}
 \displaystyle
      &\mathbb{P}_\theta(Y_{ij,t} = \text{``+''}| \mathbf{Y}_{(-ij),t} = \mathbf{y}_{(-ij),t}, \mathbf{Y}_{t-1} = \mathbf{y}_{t-1}) = \frac{\exp \left\{ \boldsymbol{\theta}^\top \Delta_{ij}^{\tilde{y}_{ij,t}\rightarrow +}( \mathbf{y}_{(-ij),t}, \mathbf{y}_{t-1}) \right\}}{\sum_{k \in \{ +, -, 0\}}\exp \left\{ \boldsymbol{\theta}^\top\Delta_{ij}^{\tilde{y}_{ij,t}\rightarrow k}( \mathbf{y}_{(-ij),t}, \mathbf{y}_{t-1}) \right\}}.   
\end{align*}
This reformulation speeds up computation, since for most statistics the calculation of global statistics is computationally more demanding than the calculation of the {change statistics} defined in \eqref{eq:change_stats}.  
%The last part holds, since we can , e.g., write $\Delta_{ij,t}^{0\rightarrow +}(\mathbf{y}_{(-ij),t}, \mathbf{y}_{t-1}) = -\Delta_{ij,t}^{+\rightarrow 0}(\mathbf{y}_{(-ij),t}, \mathbf{y}_{t-1})$ and $\Delta_{ij,t}^{0\rightarrow +}(\mathbf{y}_{(-ij),t}, \mathbf{y}_{t-1})  -\Delta_{ij,t}^{+\rightarrow -}(\mathbf{y}_{(-ij),t}, \mathbf{y}_{t-1}) = \Delta_{ij,t}^{0\rightarrow -}(\mathbf{y}_{(-ij),t}, \mathbf{y}_{t-1})$.  %Note that any {change statistics} are additive in that, e.g., $\Delta_{ij,t}^{0\rightarrow +}(\mathbf{y}_{(-ij),t}, \mathbf{y}_{t-1}) + \Delta_{ij,t}^{+\rightarrow -}(\mathbf{y}_{(-ij),t}, \mathbf{y}_{t-1})$  
Given $M$ sampled networks, we get 
\begin{equation}
\begin{aligned}
    r(\theta,\theta_0;\mathbf{y})\approx  &(\theta - \theta_0)^\top \left(\sum_{t = 1}^T \mathbf{s}(\mathbf{y}_t, \mathbf{y}_{t-1})\right) \\&-\log\left(\frac{1}{M} \sum_{m = 1}^M\exp\left\{(\theta - \theta_0)^\top \left(\sum_{t = 1}^T \mathbf{s}(\mathbf{y}_t^{(m)}, \mathbf{y}_{t-1})\right)\right\}\right),    \label{eq:lhr_approx}
\end{aligned}
\end{equation}
as an approximation of \eqref{eq:lhr}.  However, according to standard theory of exponential families, the parameter $\theta$ maximizing \eqref{eq:lhr_approx} only exists if the sum of all observed sufficient statistics  $\sum_{t = 1}^T \mathbf{s}(\mathbf{y}_t, \mathbf{y}_{t-1})$ under $\theta_0$ is inside the convex hull spanned by the sum of the sampled sufficient statistics (see Theorem 9.13 in \citealp{Barndorff-Nielsen1978}). Since this condition does not hold for arbitrary values of $\theta_0$, we modify the partial stepping algorithm under a log-normal assumption on the sufficient statistics introduced by \citet{Hummel2012} to dynamic signed networks for finding an adequate value for $\theta_0$ (details can be found in the Supplementary Material). We seed our algorithm with $\theta_0$ maximizing the pseudo-likelihood given by \eqref{eq:conditional_prob}. To obtain estimates in the cross-sectional setting of \eqref{eq:sergm}, we can use the same procedure by setting $T=1$.
%As argued in \citet{hunter2006}, there are two aspects to the uncertainty of the converged $\hat{\theta}$. First, there is the sampling error of the ML estimator, and, second, an MCMC error is inherent to the approximation of \eqref{eq:lhr} by \eqref{eq:lhr_approx}. 

%\begin{align}
 %   \hat{\mathcal{I}}(\hat{\theta}) \approx \frac{1}{M} \sum_{m = 1}^M\left(\sum_{t = 1}^T s(\mathbf{y}_t^{(m)},\mathbf{y}_{t-1}) - \mathbf{m}_{\hat{\theta}}\right)\left(\sum_{t = 1}^T s(\mathbf{y}_t^{(m)},\mathbf{y}_{t-1}) - \mathbf{m}_{\hat{\theta}}\right)^\top, \label{eq:fisher}
%\end{align}
%where $\mathbf{m}_{\hat{\theta}} = \frac{1}{M} \sum_{m = 1}^M \sum_{t = 1}^T s(\mathbf{y}_{t}^{(m)}, \mathbf{y}_{t-1})$.
To quantify the sampling error of the estimates, we rely on the theory of exponential families stating that the Fisher information $\mathcal{I}(\theta)$ equals the variance of $\sum_{t = 1}^T \mathbf{s}(\mathbf{Y}_t, \mathbf{y}_{t-1})$ under the maximum likelihood estimate $\hat{\theta}$. We can estimate the Fisher information by again sampling networks $\mathbf{Y}^{(1)}, ...,\mathbf{Y}^{(M)}$ and calculating the empirical variance of \newline $\sum_{t = 1}^T \mathbf{s}(\mathbf{y}_t^{(m)},\mathbf{y}_{t-1})$ for $m = 1, ..., M$. Due to the employed MCMC approximation, we follow standard practice of the $\mathtt{ergm}$ and $\mathtt{coda}$ packages \citep{Handcock2008,coda} and estimate the MCMC standard error by the spectral density at frequency zero of the Markov chains of the statistics. For the final variance estimate, we sum up both types of errors.  By extending the bridge sampler introduced in \citet{hunter2006} to the SERGM for dynamic networks, we can also evaluate the AIC value of the model to carry out a model selection (see Supplementary Material).

\section{Testing Structural Balance in International Cooperation and Conflict}
\label{sec:application}

\subsection{Motivation}

We now employ the SERGM to investigate relations of cooperation and conflict in the interstate network over the years 2000-2010. %Note that we also exhibit the capabilities of the SERGM in the cross-sectional context by applying it to an additional application case consisting of a classic signed network of enmity and friendship among New Guinean Highland Tribes. 
This application speaks directly to \citet{Maoz_2007}, \citet{Lerner_2016}, and the many other studies on structural balance in international relations cited above. We focus on this period since it is the most current period for which we have comprehensive and reliable data and because 9/11 provided a structural break in international relations. We do not let $\theta$ vary over time here, but it would be reasonable to assume that 9/11 altered the dynamics of the interstate network \citep[see][]{thurner2019network}. Hence $\theta$ likely changed from before to after 9/11 and we analyze only the 2000s. One example of this phenomenon is how states cooperate on their defense and security policies after 9/11. While alliances remain important, there is nowadays relatively little change in the alliance network from one year to another as ``only a dozen new alliances have emerged since 9/11'' \citep[p.730]{Kinne_2020}. Instead, a new type of formal commitment between states, defence cooperation agreements (DCA), have become widely used throughout the 1990s and 2000s \citep[see][]{Kinne_2018,Kinne_2020}.    
To ensure that we capture interstate cooperation in a meaningful manner for the period we are interested in, we depart from previous studies of structural balance in international relations and use DCAs instead of alliances to operationalize interstate cooperation. We do so for for several reasons. 

First, as noted, the contemporary alliance network is basically static, experiencing little to no shifts over time. This is a challenge for estimation but, substantively, also severely limits the extent to which alliance relations could be affected by conflict between states. In contrast, DCAs are both initiated and terminated regularly \citep{Kinne_2018}. Second, contemporary alliances are often multilateral and strongly institutionalized, meaning that if e.g. a new member joined NATO, it would result in the creation of several new alliance ties at once, but also that terminating these alliances, which have own secretariats, headquarters, and command structures, is challenging and thus empirically rare. Alliances hence do not clearly correspond to dyadic ties and have a life of their own which restricts tie deletion. In contrast, DCAs are bilateral and not as institutionalized, making them correspond much better to positive dyadic ties which can be formed but also removed \citep{Kinne_2018}. Third, as opposed to alliances, DCAs are also signed by countries which have a policy of neutrality, thus reducing the risk that some ties are structural zeros, i.e. ineligible to be formed \citep{Kinne_2020}. And fourth, most alliances only become active during armed conflict, stipulating wartime cooperation between their members \citep{Leeds_Ritter_Mitchell_Long_2002}, but their goal is to deter enemies from instigating conflict in the first place. In other words, states' formal commitment to cooperate, as demonstrated in an alliance, becomes realized only in a fraction of cases which are those where the alliance's main goal, deterrence, has failed. In contrast, DCAs specify states' commitment to and framework for peacetime, day-to-day defence cooperation regarding activities such as joint defence policies, military exercises, the co-development of military technology, and bilateral arms transfers \citep{Kinne_2018,Kinne_2020}. DCAs therefore present a better dynamic measure of regular, bilateral defence cooperation between states for the 2000s than alliances do.

\subsection{Model Specification}

To measure cooperative, positively-signed interstate relations, we thus use the DCA data collected by \citet{Kinne_2020} and consider a tie as existent and positive if a pair of states shares at least one active DCA in year $t$. For conflictious, negatively-signed relations, we follow \citet{Maoz_2007} and \citet{Lerner_2016} by using the Militarized Interstate Dispute (MID) Data provided by \citet{Palmer2021}. MIDs are defined as ``united historical cases of conflict in which the threat, display or use of military force short of war by one member state is explicitly directed towards the government, official representatives, official forces, property, or territory of another state'' \citep[p.163]{jones1996militarized}. We consider a tie to be existent and negative in year $t$ if a pair of states has at least one MID between them. We plot the resulting interstate network, consisting of positive DCA- and negative MID-ties, in the Supplementary Material. 

To specificy a SERGM for modeling this evolving network, we first follow \citet{Maoz_2007} and \citet{Lerner_2016} by including several exogenous covariates, namely $i$'s and $j$'s political difference, military capability ratio, the difference in wealth, and geographical distance. These variables' sources %together with their mathematical formulation as sufficient statistics 
are discussed in the Supplementary Material. Stemming from \eqref{eq:tsergm}, we condition on the first year for the estimation and hence effectively model the network between 2001 and 2010.  

Regarding endogenous statistics, the SERGM includes, most importantly, the four triadic terms developed above to capture the network's tendency towards or against structural balance. Theoretically, we would expect the coefficients concerning $GWESE^+(\mathbf{y}_t$) and $GWESF^+(\mathbf{y}_t)$ but not $GWESF^-(\mathbf{y}_t)$ to have positive and statistically significant coefficients. For $GWESE^-(\mathbf{y}_t)$, the expectation depends on whether we believe the state system to consist of two or of more groups \citep{Davis_1967}. The latter appears more likely for the 2000s and we may thus expect to observe a positive coefficient. Furthermore, we include the positive and negative degree statistics, to capture highly active nodes' propensity to (not) form more ties, and statistics that count the number of positive and negative edges as well as how many isolate nodes exist in each part of the network. Finally, stability terms are included to capture positive and negative ties remaining from the previous period. We term this specification Model 1 and present the results on the left side of Table \ref{tbl:res_cow}. 

We further compare Model 1 to a model specification where we replace the endogenous terms of structural balance, as depicted in Figure \ref{fig:suff_stats}, with the exogenous versions used by \citet{Maoz_2007} and \citet{Lerner_2016}, stated in \eqref{eq:stat_fof}, where $i$'s and $j$'s ties with $h$ are observed not contemporaneously but in $t-1$. We denote the corresponding statistics by $CF^+(\mathbf{y}_{t}, \mathbf{y}_{t-1})$ and $CF^-(\mathbf{y}_{t}, \mathbf{y}_{t-1})$ to quantify the effect of common friends on positive and negative ties, while the number of common enemies are $CE^+(\mathbf{y}_{t}, \mathbf{y}_{t-1})$ and $CE^-(\mathbf{y}_{t}, \mathbf{y}_{t-1})$. Each of these exogenous measures corresponds to one of our triadic endogenous statistics, e.g. $CF^+(\mathbf{y}_{t}, \mathbf{y}_{t-1})$ to $GWESF^+(\mathbf{y}_{t})$ and $CE^-(\mathbf{y}_{t}, \mathbf{y}_{t-1})$ to $GWESE^-(\mathbf{y}_{t})$. Otherwise, the two models are identical as Model 2 includes the other endogenous statistics specified in Model 1. We can thus adjudicate whether operationalizing structural balance dynamics in an endogenous manner, implying that they occur instantaneously, is preferable over the exogenous specification where these dynamics occur with a one-period time delay.  

%We estimate this model using \textcolor{red}{the stepping algorithm developed in section \ref{sec:estimation}} and present its results on the left side of table \ref{tbl:res_cow}.  

%The first one is the network of New Guinean Highland Tribes which are connected either by friendship or enmity, this is a standard dataset used to illustrate methods for signed networks. The second application is the network of alliances and conflict between states in the period 1866-1870, as the European powers realigned their international relations following the Austro-Prussian War. The Highland Tribes network is a simple first example application as it is only observed once and includes no exogenous covariates. The interstate network is closer to applications substantive researchers will be interested in as it is observed over time, influenced by available covariates, and speaks directly to \citet{Maoz_2007} or \citet{Lerner_2016}. Descriptive information on these two datasets is presented in the Supplementary Material.

\subsection{Results}
\begin{table}[!tbp]
\caption{Estimated coefficients and confidence intervals of the two model specifications detailed above. Dashes indicate the exclusion of covariates in a model specification. $\Delta$AIC indicates the difference between the AIC values of Model 1 and 2. \label{tbl:res_cow}} 
\begin{center}
\begin{tabular}{lccccc}
\hline
\multicolumn{1}{l}{\  }&\multicolumn{2}{c}{\  Model 1}&\multicolumn{1}{c}{\  }&\multicolumn{2}{c}{\  Model 2}\tabularnewline
\cline{2-3} \cline{5-6}
\multicolumn{1}{l}{}& \multicolumn{1}{c}{Coef.}&\multicolumn{1}{c}{CI}&\multicolumn{1}{c}{}&\multicolumn{1}{c}{Coef.}&\multicolumn{1}{c}{CI}\tabularnewline
\hline
Edges $+$&   -1.161&[-1.59,-0.732]&&-0.689&[-1.203,-0.175]\tabularnewline
Edges $-$&-1.754&[-2.142,-1.366]&&-1.469&[-1.912,-1.026]\tabularnewline
Isolates $+$&0.667&[-0.203,1.537]&&0.462&[-0.422,1.346]\tabularnewline
Isolates $-$&-1.188&[-2.319,-0.057]&&-0.474&[-1.617,0.669]\tabularnewline
Stability $+$&7.447&[7.331,7.563]&&7.502&[7.379,7.625]\tabularnewline
Stability $-$&5.531&[5.262,5.8]&&5.594&[5.306,5.882]\tabularnewline
Abs. Polity Diff. $+$&-0.022&[-0.032,-0.012]&&-0.017&[-0.027,-0.007]\tabularnewline
Abs. Polity Diff. $-$&0.004&[-0.016,0.024]&&0.012&[-0.01,0.034]\tabularnewline
CINC Ratio $+$&0.186&[0.117,0.255]&&0.202&[0.129,0.275]\tabularnewline
CINC Ratio $-$&-0.168&[-0.293,-0.043]&&-0.14&[-0.279,-0.001]\tabularnewline
Abs. GDP Diff. $+$&-0.521&[-0.57,-0.472]&&-0.495&[-0.554,-0.436]\tabularnewline
Abs. GDP Diff. $-$&-1.04&[-2.651,0.571]&&-1.311&[-3.069,0.447]\tabularnewline
Abs. Distance $\pm$&3.324&[0.515,6.133]&&2.796&[-0.428,6.02]\tabularnewline
$GWESE^+$ (Fig. \ref{fig:suff_stats}a)&0.618&[0.308,0.928]&&-&\tabularnewline
$GWESE^-$ (Fig. \ref{fig:suff_stats}b)&0.515&[0.199,0.831]&&-&\tabularnewline
$GWESF^+$ (Fig. \ref{fig:suff_stats}c)&0.489&[0.415,0.563]&&-&\tabularnewline
$GWESF^-$ (Fig. \ref{fig:suff_stats}d)&0.319&[0.178,0.46]&&-&\tabularnewline
$GWD^+$&-2.214&[-2.577,-1.851]&&-2.625&[-3.015,-2.235]\tabularnewline
$GWD^-$&-0.321&[-1.617,0.975]&&-0.998&[-2.276,0.28]\tabularnewline
$CF^+$&-&&&0.069&[0.051,0.087]\tabularnewline
$CF^-$&-&&&0.077&[0.04,0.114]\tabularnewline
$CE^+$&-&&&0.374&[-0.042,0.79]\tabularnewline
$CE^-$&-&&&0.304&[-0.239,0.847]\tabularnewline
\hline
$\Delta$AIC& 0 &&&599.894&\tabularnewline
\hline
\end{tabular}\end{center}
\end{table}

Below, we interpret the results of the endogenous network terms and their exogenous equivalents. We discuss the coefficient estimates of the exogenous covariates in the Supplementary Material. As expected, both the $GWESF^+(\mathbf{y}_t)$ and the $GWESE^+(\mathbf{y}_t)$ terms exhibit positive and statistically significant coefficients, with neither confidence interval encompassing zero. These results align with structural balance theory in that both ``the friend of my friend'' and ``the enemy of my enemy'' are my friend. But we also find that the $GWESF^-(\mathbf{y}_t)$ and  $GWESE^-(\mathbf{y}_t)$ coefficients are positive and statistically significant, albeit with smaller effects and confidence intervals closer to zero than in the case of the first two statistics. In the studied interstate network, there is thus also a tendency towards enemies of enemies being enemies. This echoes the point that triangles with three negative ties are imbalanced only in systems with two subsets \citep{Davis_1967}, a condition that may have been present in the highly bipolar first half of the Cold War, but not more than a decade after its termination. This result is thus consistent with the verdict that, against early formulations of structural balance theory \citep{Heider_1946,Cartwright_Harary_1956}, ``if two negative relations are given, balance can be obtained either when the third relationship is positive or when it is negative'' \citep[p.206]{heider1958psychology}. Observing that the effect of $GWESE^-(\mathbf{y}_t)$ is positive and statistically significant underlines the importance of overall network structure for the predictions of structural balance theory.

We also find that friends of friends have an increased probability of being enemies. In the international relations of the 2000s, what seems to hold is that both enemies of enemies and friends of friends are more likely to interact than if they did not share relations with a common third state. Friends of friends being more likely to fight than to have no relation at all suggests that shared relations may also indicate the ``reachability'' of one state to another within a system where some dyads, e.g., that between Lesotho and Belize, have a very low structural probability of ever being active \citep[see, e.g.][]{quackenbush2006identifying}. Triadic closure, regardless of the sign, thus exists also in the network of cooperation and conflict between states. However, we observe that the tendency towards such closure is stronger for structurally balanced relations than for structurally imbalanced ones.

A comparison of the two model specifications shown in Table \ref{tbl:res_cow} allows us to ascertain whether specifying the triadic relationships endogenously affects substantive results and model performance. Here, it is visible that the AIC of the model with the endogenous statistics is lower than that with their exogenous versions. Specifying interdependent dynamics in the interstate network via endogenous covariates hence increases model performance compared to trying to capture them by including lagged, exogenous variables. 

%GWESF $+$ and GWESE $+$ both exhibit positive and statistically significant coefficients, thus offering support for the idea that both friends of friends and enemies of enemies are ultimately friends.
More strikingly, Table \ref{tbl:res_cow} shows that the substantive results of the corresponding endogenous and exogenous measures of structural balance dynamics differ significantly. Contrasting the results under the endogenous and exogenous model specification, the latter offers much more limited support for these notions. While the coefficient of $CF^+(\mathbf{y}_t, \mathbf{y}_{t-1})$ is positive and statistically significant, its effect size is still very close to zero. The ``friends of friends are friends''-effect is thus found to be substantively negligible in Model 2. In contrast, the coefficient of $CE^+(\mathbf{y}_t, \mathbf{y}_{t-1})$ is positive and substantively larger, while its 95$\%$-confidence interval includes zero, indicating that the model cannot statistically distinguish it from zero as its estimation is very imprecise. The statistics $CF^-(\mathbf{y}_t, \mathbf{y}_{t-1})$ and $CE^-(\mathbf{y}_t, \mathbf{y}_{t-1})$ mirror their corresponding endogenous terms from Model 1 in that both exhibit positive coefficients but, again, the first is substantively much smaller and the second one very imprecisely estimated. On the whole, this comparison of an endogenous and an exogenous specification of the triadic configurations motivated by structural balance theory thus shows that Model 1 is preferable over Model 2. The model including endogenous terms thus not only provides better performance than that with their exogenous counterparts but these terms are also estimated to be more influential and more precisely.

\subsection{Model Assessment}

\begin{figure}[t!]\centering
	 \centering
    \includegraphics[width=0.7\linewidth]{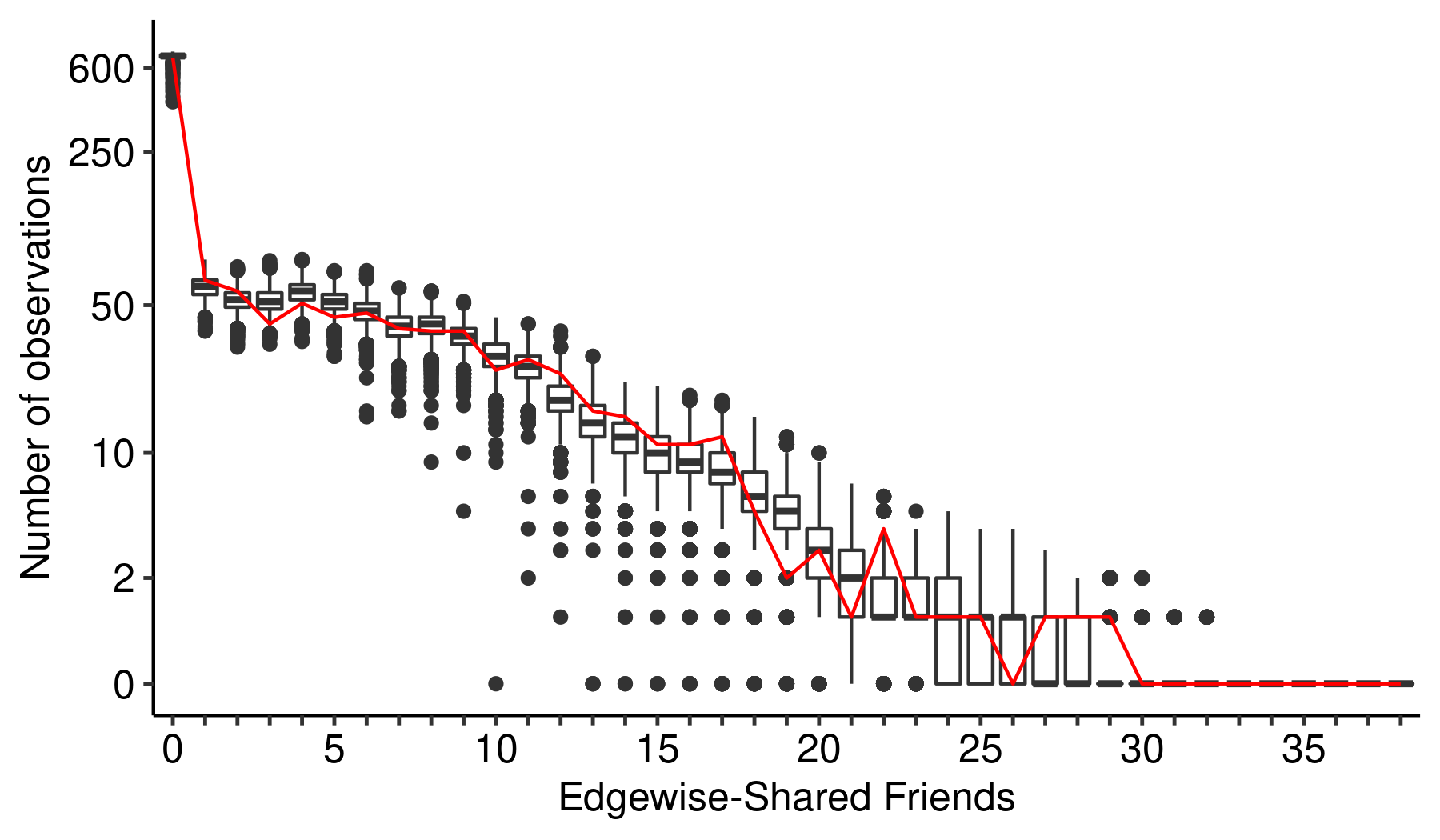}
    \includegraphics[width=0.7\linewidth]{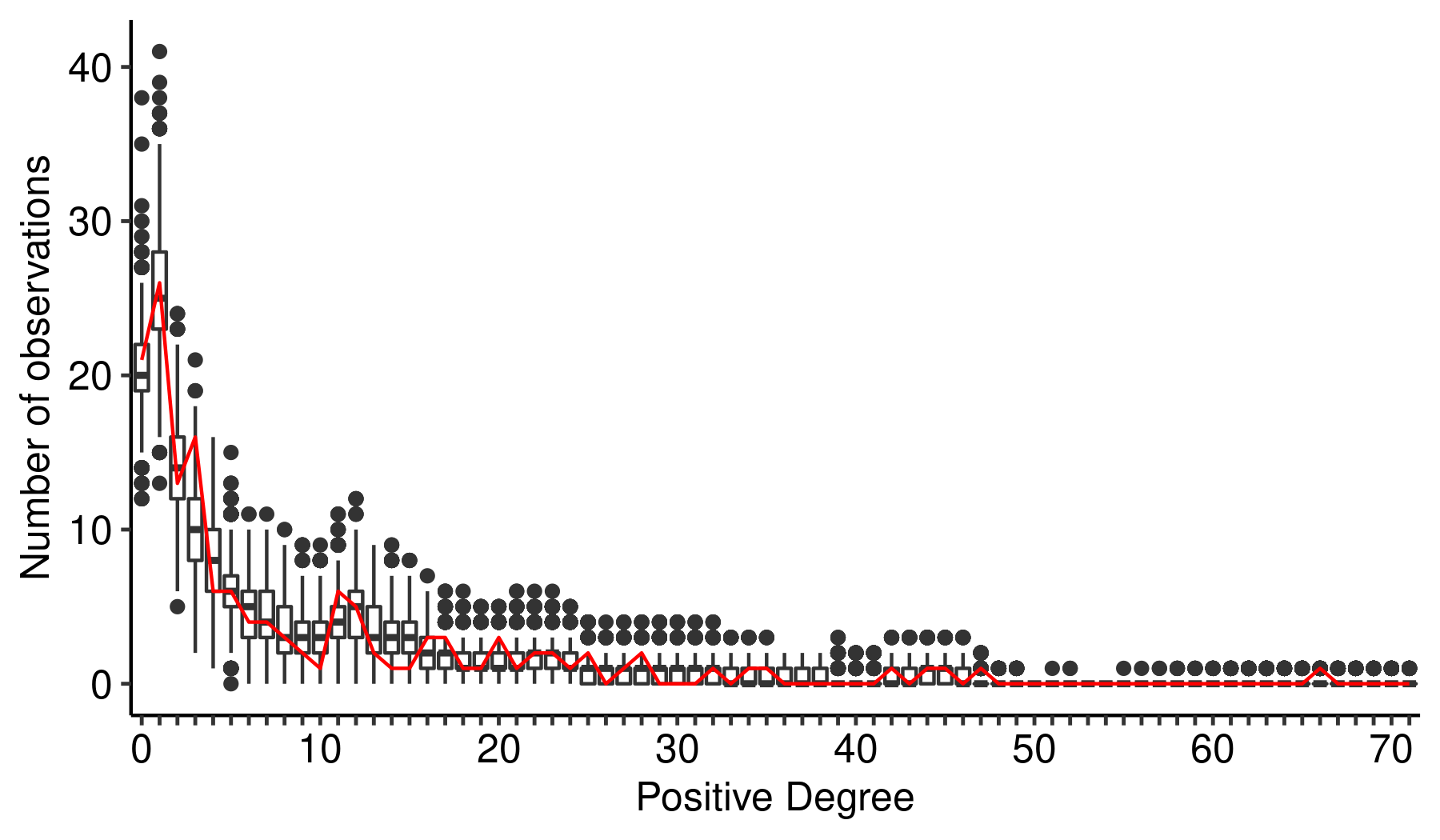}
    \includegraphics[width=0.35\linewidth]{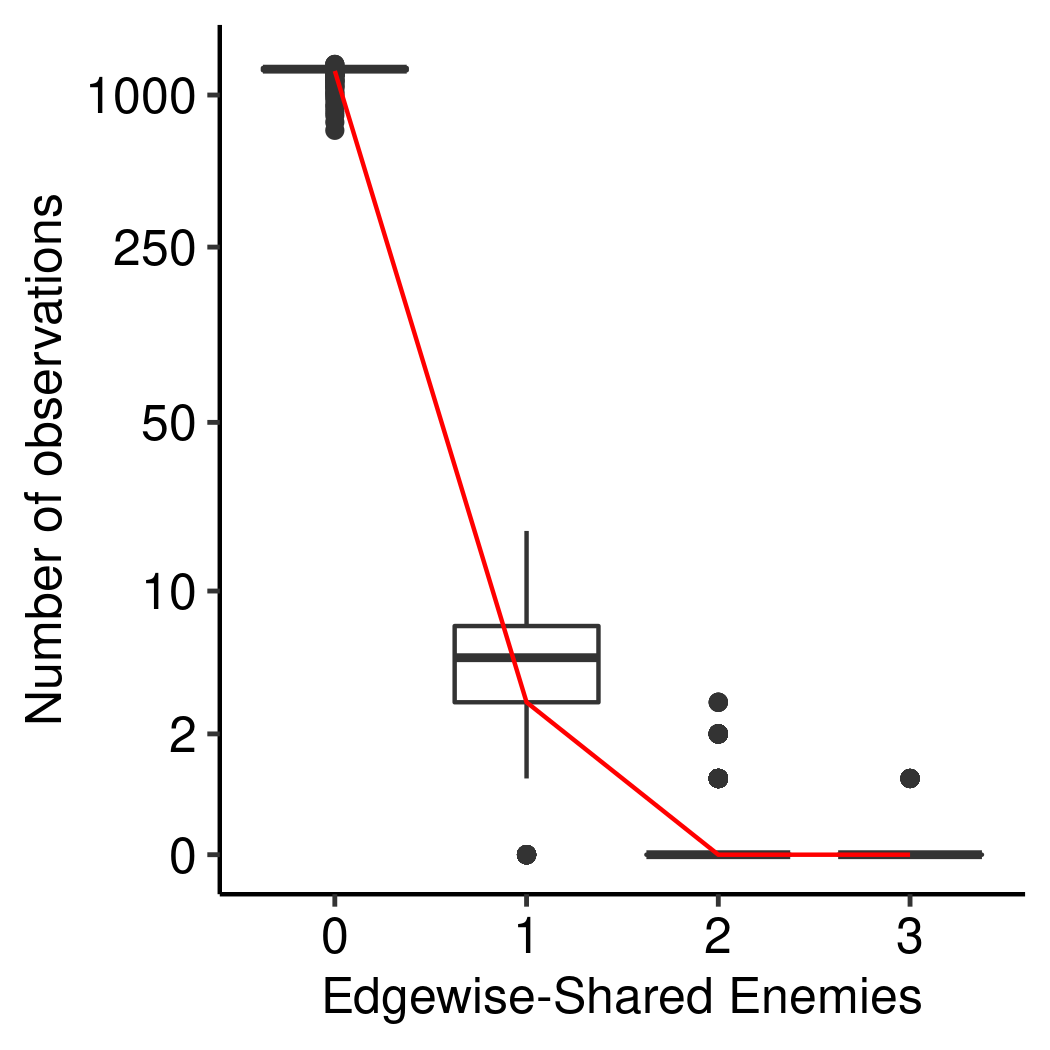}
   \includegraphics[width=0.35\linewidth]{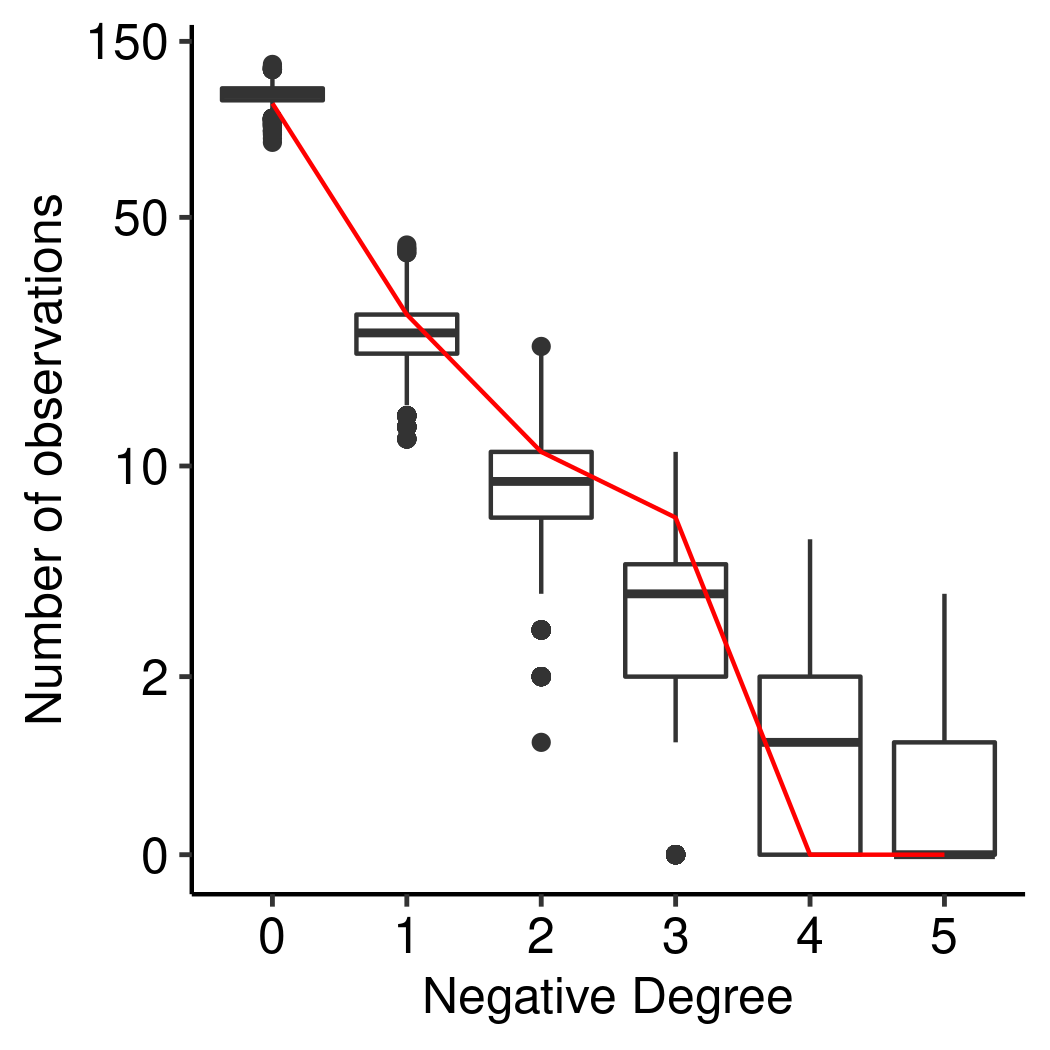}
	\caption{Goodness-of-fit assessment in year 2005.}
	\label{fig:gof}
\end{figure}

To assess the fit of the estimated SERGM, we employ a graphical tool inspired by \citet{Hunter2008b} to evaluate whether it can adequately represent topologies of the observed network not explicitly incorporated as sufficient statistics in \eqref{eq:tsergm}. Therefore, we sample networks from \eqref{eq:tsergm}, compute the statistics, summarize them, and then compare this summary to the statistics evaluated on the observed network. Heuristically, a model generating simulations that better reflect the observed values also has a better goodness-of-fit. To cover signed networks, we investigate the observed and simulated distributions of positive and negative degrees, and edgewise-shared enemies and friends in the interstate network.
%\citet{Hunter2008b} proposed this technique of goodness-of-fit plots for binary networks, we extend it to cover signed networks.

We report the goodness-of-fit plots for Model 1 from Table \ref{tbl:res_cow} in Figure \ref{fig:gof} for the year 2005.  In each subplot, a series of box plots display the distribution of a given value of the statistic under consideration over the networks simulated from the model via the Gibbs Sampler detailed in Section \ref{sec:estimation}. The red line indicates where the statistic is measured in the observed network and should thus, ideally, lie close to the median value of the simulated networks, i.e., the center of the box plots. In Figure \ref{fig:gof}, this is the case for all four statistics, indicating that Model 1 under the estimated parameters generalizes well to network topologies not explicitly incorporated in the sufficient statistics.    

Together, the results presented here indicate that the SERGM is able to uncover structural balance dynamics in the interstate network and is preferable over approaches that seek to model signed interstate networks under conditional independence, but also that further substantial research on structural balance in international relations is neeeded. The Supplementary Material employs the SERGM to analyze a cross-sectional network, representing enmity and friendship among New Guinean Highland Tribes \citep{Hage_Harary_1984}, and shows its applicability when there is no observable temporal dependence structure.

\section{Discussion}

We extended the core regression model for network data to dynamic and cross-sectional signed networks. Given the theoretical foundation of structural balance, we introduce novel endogenous statistics that offer better performance than operationalizing them by lagged covariates, as commonly done in previous research. Finally, we apply the method to recent data on militarized interstate disputes and defense cooperation agreements and provide a software implementation with the $\mathtt{R}$ package $\mathtt{ergm.sign}$. 

From a substantive point of view, this research offers new insights on the empirical testing of structural balance theory and challenges earlier inferential studies on the topic. How one captures structural balance matters. We show that an approach relying on past observations of some ties within a triad to measure structural balance as an exogenous variable can mischaracterize triadic (im-)balance. We thus develop endogenous balance measures that can be used in the SERGM framework and show empirically that these endogenous measures result in different substantive results as well as increased model performance as compared to the exogenous ones. Most importantly, the exogenous measures do not affect tie formation consistent with structural balance theory, whereas when employing the endogenous ones, we find evidence in line with it. States are thus more likely to cooperate if they share common partners or are hostile to the same enemies. This indicates that there is structural balance in interstate cooperation and conflict, at least when studying the 2000s. Future work in International Relations should seek to build on this fundamental result to test whether it also holds for earlier periods, for instance the bipolar Cold War years, and how structural balance interacts with exogenous factors such as military capabilities. Beyond International Relations, the SERGM will also serve to advance research across all Social Sciences, allowing researchers to investigate tie formation in networks of friendship and enmity between school children, gangs, or social media accounts.      

At the same time, we find that, generally, states appear more likely to interact, positively or negatively, when they share friends or enemies. Substantively, this result suggests that, additional to structural balance, something else is at play and may indicate that some state dyads are structurally very unlikely to ever be active, due to the countries' distance, lack of economic development, and/or power projection capabilities, mirroring research on politically ``relevant'' or ``active'' dyads \citep[see][]{quackenbush2006identifying}. But this implied variation in ``reachability'' between states also points to the fact that structural balance theory was developed on complete networks, where every possible ties is realized with either a negative or a positive sign, while empirical networks are usually incomplete \citep[see][ch.5]{Easley_Kleinberg_2010}. It thus lends some support to \citeauthor{Lerner_2016}'s \citeyearpar{Lerner_2016} argument that tests of structural balance theory should not examine states' marginal probability to cooperate or fight, but instead their probability of cooperating or fighting \textsl{conditional upon them interacting}. However, following \citeauthor{Lerner_2016}'s \citeyearpar[Sec. 4.2.1]{Lerner_2016} argument on the use of ERGMs in conjunction with this conditional viewpoint, it becomes evident that \eqref{eq:tsergm} is consistent with it. Defining $\mathbf{Y}^{|\pm|}$ with $Y^{|\pm|}_{ij} = 1 \text{ if } Y_{ij} \neq$ ``0''  as the random adjacency matrix describing any type of interaction and $\mathcal{Y}$, be it positive or negative, one can derive the following conditional probability distribution
\begin{align}\label{eq:tsergm_cond}
\mathbb{P}_{\boldsymbol{\theta}}(\mathbf{Y}_t = \mathbf{y}_t|\mathbf{Y}^{|\pm|}_t = \mathbf{y}^{|\pm|}_t,\mathbf{Y}_{t-1} =  \mathbf{y}_{t-1}) = \frac{\exp \left\{ \boldsymbol{\theta}^\top\mathbf{s}( \mathbf{y}_{t}, \mathbf{y}_{t-1}) \right\}}{\tilde{\kappa}(\boldsymbol{\theta}, \mathbf{y}_{t-1},\mathbf{y}^{|\pm|}_t)} ~ \forall ~ \mathbf{y}_t \in \mathcal{Y}^{\pm}, 
\end{align} 
where $\tilde{\kappa}(\boldsymbol{\theta}, \mathbf{y}_{t-1},\mathbf{y}^{|\pm|}_t) = \sum_{\tilde{\mathbf{y}} \in  \mathcal{Y}^{\pm}} \mathbb{I}(\mathbf{\tilde{y}}^{|\pm|} = \mathbf{y}^{|\pm|}_t)\exp \left\{ \boldsymbol{\theta}^\top\mathbf{s}( \tilde{\mathbf{y}}, \mathbf{y}_{t-1}) \right\}$. The conditional distribution \eqref{eq:tsergm_cond} is thus a SERGM with support limited to networks where $\mathbf{y}^{|\pm|}_t$ is equal to the observed network and the coefficients of \eqref{eq:tsergm_cond} are unchanged. Therefore, \eqref{eq:tsergm} implies \eqref{eq:tsergm_cond}. 

%\textcolor{red}{How can we make the model work in the constrained lerner setting, does this work? Or does it even work generally?}

Alternatively, some dyads' lack of ``reachability'' may also indicate that dependency structures are not fully global, even in international relations where all actors know of each other. Major powers should generally be able to reach all other states in the system, thus also making their actions globally relevant, but smaller countries' reach and relevance will be more locally limited. Since the more general framework of ERGMs in \eqref{eq:ergm} relies on homogeneity assumptions implying that each endogenous mechanism has the same effect in the entire network, model \eqref{eq:tsergm} might assume dependence between relations where, in reality, there is none. One possible endeavor for future research would be adapting local dependence \citep{Schweinberger2015} to signed and dynamic networks. This approach assumes complex dependency solely within either observed or unobserved groups of the actors, solving the obstacle of ``reachability'' between some countries in the network.  At the same time, other extensions of ERGMs, be it actor-specific random effects or curved ERGMs where $\alpha$ in \eqref{eq:stat_gwese} is estimated from the data, are also feasible  under \eqref{eq:sergm} and \eqref{eq:tsergm}.

\newpage
\bibliographystyle{chicago}
\bibliography{references}

\begin{thebibliography}{}

\bibitem[\protect\citeauthoryear{Arinik, Figueiredo, and Labatut}{Arinik
  et~al.}{2020}]{Arinik_2020}
Arinik, N., R.~Figueiredo, and V.~Labatut (2020).
\newblock Multiple partitioning of multiplex signed networks: Application to
  european parliament votes.
\newblock {\em Social Networks\/}~{\em 60}, 83–102.

\bibitem[\protect\citeauthoryear{Barndorff-Nielsen}{Barndorff-Nielsen}{1978}]{Barndorff-Nielsen1978}
Barndorff-Nielsen, O. (1978).
\newblock {\em {Information and exponential families in statistical theory}}.
\newblock New York: Wiley.

\bibitem[\protect\citeauthoryear{Bonacich and Lloyd}{Bonacich and
  Lloyd}{2004}]{Bonacich2004}
Bonacich, P. and P.~Lloyd (2004).
\newblock {Calculating status with negative relations}.
\newblock {\em Social Networks\/}~{\em 26\/}(4), 331--338.

\bibitem[\protect\citeauthoryear{Bramson, Hoefman, Schoors, and
  Ryckebusch}{Bramson et~al.}{2021}]{Bramson_Hoefman_Schoors_Ryckebusch_2021}
Bramson, A., K.~Hoefman, K.~Schoors, and J.~Ryckebusch (2021).
\newblock Diplomatic relations in a virtual world.
\newblock {\em Political Analysis\/}, 1–22.

\bibitem[\protect\citeauthoryear{Cartwright and Harary}{Cartwright and
  Harary}{1956}]{Cartwright_Harary_1956}
Cartwright, D. and F.~Harary (1956).
\newblock Structural balance: A generalization of heider’s theory.
\newblock {\em Psychological Review\/}~{\em 63\/}(5), 277–293.

\bibitem[\protect\citeauthoryear{Davis}{Davis}{1967}]{Davis_1967}
Davis, J.~A. (1967).
\newblock Clustering and structural balance in graphs.
\newblock {\em Human Relations\/}~{\em 20\/}(2), 181–187.

\bibitem[\protect\citeauthoryear{De~Nooy and Kleinnijenhuis}{De~Nooy and
  Kleinnijenhuis}{2013}]{DeNooy_Kleinnijenhuis_2013}
De~Nooy, W. and J.~Kleinnijenhuis (2013).
\newblock Polarization in the media during an election campaign: A dynamic
  network model predicting support and attack among political actors.
\newblock {\em Political Communication\/}~{\em 30\/}(1), 117–138.

\bibitem[\protect\citeauthoryear{Doreian and Mrvar}{Doreian and
  Mrvar}{2009}]{Doreian2009}
Doreian, P. and A.~Mrvar (2009).
\newblock {Partitioning signed social networks}.
\newblock {\em Social Networks\/}~{\em 31\/}(1), 1--11.

\bibitem[\protect\citeauthoryear{Doreian and Mrvar}{Doreian and
  Mrvar}{2015}]{Doreian_Mrvar_2015}
Doreian, P. and A.~Mrvar (2015).
\newblock Structural balance and signed international relations.
\newblock {\em Journal of Social Structure\/}~{\em 16\/}(1), 1–49.

\bibitem[\protect\citeauthoryear{Easley and Kleinberg}{Easley and
  Kleinberg}{2010}]{Easley_Kleinberg_2010}
Easley, D. and J.~Kleinberg (2010).
\newblock {\em Networks, crowds, and markets: Reasoning about a highly
  connected world}.
\newblock Cambridge University Press.

\bibitem[\protect\citeauthoryear{Everett and Borgatti}{Everett and
  Borgatti}{2014}]{Everett2014}
Everett, M.~G. and S.~P. Borgatti (2014).
\newblock {Networks containing negative ties}.
\newblock {\em Social Networks\/}~{\em 38\/}(1), 111--120.

\bibitem[\protect\citeauthoryear{Frank and Strauss}{Frank and
  Strauss}{1986}]{frank1986}
Frank, O. and D.~Strauss (1986).
\newblock {Markov graphs}.
\newblock {\em Journal of the American Statistical Association\/}~{\em
  81\/}(395), 832--842.

\bibitem[\protect\citeauthoryear{Hage and Harary}{Hage and
  Harary}{1984}]{Hage_Harary_1984}
Hage, P. and F.~Harary (1984).
\newblock {\em Structural Models in Anthropology}.
\newblock Cambridge: Cambridge University Press.

\bibitem[\protect\citeauthoryear{Handcock}{Handcock}{2003}]{handcock2003assessing}
Handcock, M. (2003).
\newblock {Assessing degeneracy in statistical models of social networks}.
\newblock Technical report, University of Washington.

\bibitem[\protect\citeauthoryear{Handcock, Hunter, Butts, Goodreau, and
  Morris}{Handcock et~al.}{2008}]{Handcock2008}
Handcock, M.~S., D.~R. Hunter, C.~T. Butts, S.~M. Goodreau, and M.~Morris
  (2008).
\newblock {statnet: Software tools for the representation, visualization,
  analysis and simulation of network data}.
\newblock {\em Journal of Statistical Software\/}~{\em 24\/}(1), 1--11.

\bibitem[\protect\citeauthoryear{Harary}{Harary}{1961}]{Harary_1961}
Harary, F. (1961).
\newblock A structural analysis of the situation in the middle east in 1956.
\newblock {\em Journal of Conflict Resolution\/}~{\em 5\/}(2), 167–178.

\bibitem[\protect\citeauthoryear{Hart}{Hart}{1974}]{Hart_1974}
Hart, J. (1974).
\newblock Symmetry and polarization in the european international system, 1870
  - 1879: A methodological study.
\newblock {\em Journal of Peace Research\/}~{\em 11\/}(3), 229–244.

\bibitem[\protect\citeauthoryear{Healy and Stein}{Healy and
  Stein}{1973}]{Healy_Stein_1973}
Healy, B. and A.~Stein (1973).
\newblock The balance of power in international history: Theory and reality.
\newblock {\em Journal of Conflict Resolution\/}~{\em 17\/}(1), 33–61.

\bibitem[\protect\citeauthoryear{Heider}{Heider}{1946}]{Heider_1946}
Heider, F. (1946).
\newblock Attitudes and cognitive organization.
\newblock {\em Journal of Psychology\/}~{\em 21\/}(1), 107–112.

\bibitem[\protect\citeauthoryear{Heider}{Heider}{1958}]{heider1958psychology}
Heider, F. (1958).
\newblock {\em The psychology of interpersonal relations.}
\newblock New York: Wiley.

\bibitem[\protect\citeauthoryear{Holland and Leinhardt}{Holland and
  Leinhardt}{1972}]{Holland1972}
Holland, P.~W. and S.~Leinhardt (1972).
\newblock {Some evidence on the transitivity of positive interpersonal
  sentiment}.
\newblock {\em American Journal of Sociology\/}~{\em 77\/}(6), 1205--1209.

\bibitem[\protect\citeauthoryear{Huitsing, Snijders, Van~Duijn, and
  Veenstra}{Huitsing et~al.}{2014}]{Huitsing_2014}
Huitsing, G., T.~A.~B. Snijders, M.~A.~J. Van~Duijn, and R.~Veenstra (2014).
\newblock Victims, bullies, and their defenders: A longitudinal study of the
  coevolution of positive and negative networks.
\newblock {\em Development and Psychopathology\/}~{\em 26\/}(3), 645–659.

\bibitem[\protect\citeauthoryear{Huitsing, van Duijn, Snijders, Wang, Sainio,
  Salmivalli, and Veenstra}{Huitsing et~al.}{2012}]{Huitsing_2012}
Huitsing, G., M.~A. van Duijn, T.~A. Snijders, P.~Wang, M.~Sainio,
  C.~Salmivalli, and R.~Veenstra (2012).
\newblock Univariate and multivariate models of positive and negative networks:
  Liking, disliking, and bully–victim relationships.
\newblock {\em Social Networks\/}~{\em 34\/}(4), 645–657.

\bibitem[\protect\citeauthoryear{Hummel, Hunter, and Handcock}{Hummel
  et~al.}{2012}]{Hummel2012}
Hummel, R.~M., D.~R. Hunter, and M.~S. Handcock (2012).
\newblock {Improving simulation-based algorithms for fitting ERGMs}.
\newblock {\em Journal of Computational and Graphical Statistics\/}~{\em
  21\/}(4), 920--939.

\bibitem[\protect\citeauthoryear{Hunter}{Hunter}{2007}]{hunter2007curved}
Hunter, D.~R. (2007).
\newblock {Curved exponential family models for social networks}.
\newblock {\em Social Networks\/}~{\em 29\/}(2), 216--230.

\bibitem[\protect\citeauthoryear{Hunter, Goodreau, and Handcock}{Hunter
  et~al.}{2008}]{Hunter2008b}
Hunter, D.~R., S.~M. Goodreau, and M.~S. Handcock (2008).
\newblock {Goodness of fit of social network models}.
\newblock {\em Journal of the American Statistical Association\/}~{\em
  103\/}(481), 248--258.

\bibitem[\protect\citeauthoryear{Hunter and Handcock}{Hunter and
  Handcock}{2006}]{hunter2006}
Hunter, D.~R. and M.~S. Handcock (2006).
\newblock {Inference in curved exponential family models for networks}.
\newblock {\em Journal of Computational and Graphical Statistics\/}~{\em
  15\/}(3), 565--583.

\bibitem[\protect\citeauthoryear{Hunter, Handcock, Butts, Goodreau, and
  Morris}{Hunter et~al.}{2008}]{Hunter2008}
Hunter, D.~R., M.~S. Handcock, C.~T. Butts, S.~M. Goodreau, and M.~Morris
  (2008).
\newblock {ergm: A package to fit, simulate and diagnose exponential-family
  models for networks}.
\newblock {\em Journal of Statistical Software\/}~{\em 24\/}(3), 1--29.

\bibitem[\protect\citeauthoryear{Jiang}{Jiang}{2015}]{Jiang2015}
Jiang, J.~Q. (2015).
\newblock {Stochastic block model and exploratory analysis in signed networks}.
\newblock {\em Physical Review E\/}~{\em 91\/}(6), 062805.

\bibitem[\protect\citeauthoryear{Jones, Bremer, and Singer}{Jones
  et~al.}{1996}]{jones1996militarized}
Jones, D.~M., S.~A. Bremer, and J.~D. Singer (1996).
\newblock Militarized interstate disputes, 1816--1992: Rationale, coding rules,
  and empirical patterns.
\newblock {\em Conflict Management and Peace Science\/}~{\em 15\/}(2),
  163--213.

\bibitem[\protect\citeauthoryear{Kinne}{Kinne}{2018}]{Kinne_2018}
Kinne, B.~J. (2018).
\newblock Defense cooperation agreements and the emergence of a global security
  network.
\newblock {\em International Organization\/}~{\em 72\/}(4), 799–837.

\bibitem[\protect\citeauthoryear{Kinne}{Kinne}{2020}]{Kinne_2020}
Kinne, B.~J. (2020).
\newblock The defense cooperation agreement dataset (dcad).
\newblock {\em Journal of Conflict Resolution\/}~{\em 64\/}(4), 729–755.

\bibitem[\protect\citeauthoryear{Kinne and Maoz}{Kinne and
  Maoz}{2022}]{Kinne_Maoz_2022}
Kinne, B.~J. and Z.~Maoz (2022).
\newblock Local politics, global consequences: How structural imbalance in
  domestic political networks aﬀects international relations.
\newblock {\em Journal of Politics\/}~{\em (OnlineFirst)}.

\bibitem[\protect\citeauthoryear{Lee, Muncaster, and Zinnes}{Lee
  et~al.}{1994}]{Lee_Muncaster_Zinnes_1994}
Lee, S.~C., R.~G. Muncaster, and D.~A. Zinnes (1994).
\newblock The friend of my enemy is my enemy: Modeling triadic internation
  relationships.
\newblock {\em Synthese\/}~{\em 100\/}(3), 333–358.

\bibitem[\protect\citeauthoryear{Leeds, Ritter, Mitchell, and Long}{Leeds
  et~al.}{2002}]{Leeds_Ritter_Mitchell_Long_2002}
Leeds, B., J.~Ritter, S.~Mitchell, and A.~Long (2002).
\newblock Alliance treaty obligations and provisions, 1815-1944.
\newblock {\em International Interactions\/}~{\em 28\/}(3), 237–260.

\bibitem[\protect\citeauthoryear{Lerner}{Lerner}{2016}]{Lerner_2016}
Lerner, J. (2016).
\newblock Structural balance in signed networks: Separating the probability to
  interact from the tendency to fight.
\newblock {\em Social Networks\/}~{\em 45}, 66–77.

\bibitem[\protect\citeauthoryear{Leskovec, Huttenlocher, and
  Kleinberg}{Leskovec et~al.}{2010}]{Leskovec_2010}
Leskovec, J., D.~Huttenlocher, and J.~Kleinberg (2010).
\newblock Signed networks in social media.
\newblock In {\em Proceedings of the 28th international conference on Human
  factors in computing systems}, pp.\  1361--1370.

\bibitem[\protect\citeauthoryear{Lusher, Koskinen, and Robins}{Lusher
  et~al.}{2012}]{Lusher2012}
Lusher, D., J.~Koskinen, and G.~Robins (2012).
\newblock {\em {Exponential random graph models for social networks}}.
\newblock Cambridge: Cambridge University Press.

\bibitem[\protect\citeauthoryear{Maoz, Terris, Kuperman, and Talmud}{Maoz
  et~al.}{2007}]{Maoz_2007}
Maoz, Z., L.~G. Terris, R.~D. Kuperman, and I.~Talmud (2007).
\newblock What is the enemy of my enemy? causes and consequences of imbalanced
  international relations, 1816–2001.
\newblock {\em Journal of Politics\/}~{\em 69\/}(1), 100–115.

\bibitem[\protect\citeauthoryear{McDonald and Rosecrance}{McDonald and
  Rosecrance}{1985}]{McDonald_Rosecrance_1985}
McDonald, H.~B. and R.~Rosecrance (1985).
\newblock Alliance and structural balance in the international system: A
  reinterpretation.
\newblock {\em Journal of Conflict Resolution\/}~{\em 29\/}(1), 57–82.

\bibitem[\protect\citeauthoryear{Nakamura, Tita, and Krackhardt}{Nakamura
  et~al.}{2020}]{Nakamura_Tita_Krackhardt_2020}
Nakamura, K., G.~Tita, and D.~Krackhardt (2020).
\newblock Violence in the “balance”: A structural analysis of how rivals,
  allies, and third-parties shape inter-gang violence.
\newblock {\em Global Crime\/}~{\em 21\/}(1), 3–27.

\bibitem[\protect\citeauthoryear{Palmer, McManus, D'Orazio, Kenwick, Karstens,
  Bloch, Dietrich, Kahn, Ritter, and Soules}{Palmer et~al.}{2021}]{Palmer2021}
Palmer, G., R.~W. McManus, V.~D'Orazio, M.~R. Kenwick, M.~Karstens, C.~Bloch,
  N.~Dietrich, K.~Kahn, K.~Ritter, and M.~J. Soules (2021).
\newblock The mid5 dataset, 2011–2014: Procedures, coding rules, and
  description.
\newblock {\em Conflict Management and Peace Science\/}~{\em 39\/}(4),
  470--482.

\bibitem[\protect\citeauthoryear{Plummer, Best, Cowles, and Vines}{Plummer
  et~al.}{2006}]{coda}
Plummer, M., N.~Best, K.~Cowles, and K.~Vines (2006).
\newblock Coda: Convergence diagnosis and output analysis for mcmc.
\newblock {\em R News\/}~{\em 6\/}(1), 7--11.

\bibitem[\protect\citeauthoryear{Quackenbush}{Quackenbush}{2006}]{quackenbush2006identifying}
Quackenbush, S.~L. (2006).
\newblock Identifying opportunity for conflict: Politically active dyads.
\newblock {\em Conflict Management and Peace Science\/}~{\em 23\/}(1), 37--51.

\bibitem[\protect\citeauthoryear{{R Core Team}}{{R Core Team}}{2021}]{R}
{R Core Team} (2021).
\newblock {\em R: A language and environment for statistical computing}.
\newblock Vienna, Austria: R Foundation for Statistical Computing.

\bibitem[\protect\citeauthoryear{Saiz, Gómez-Gardeñes, Nuche, Girón, Pueyo,
  and Alados}{Saiz et~al.}{2017}]{Saiz_2017}
Saiz, H., J.~Gómez-Gardeñes, P.~Nuche, A.~Girón, Y.~Pueyo, and C.~L. Alados
  (2017).
\newblock Evidence of structural balance in spatial ecological networks.
\newblock {\em Ecography\/}~{\em 40\/}(6), 733–741.

\bibitem[\protect\citeauthoryear{Schweinberger}{Schweinberger}{2011}]{schweinberger2011instability}
Schweinberger, M. (2011).
\newblock {Instability, sensitivity, and degeneracy of discrete exponential
  families}.
\newblock {\em Journal of the American Statistical Association\/}~{\em
  106\/}(496), 1361--1370.

\bibitem[\protect\citeauthoryear{Schweinberger and Handcock}{Schweinberger and
  Handcock}{2015}]{Schweinberger2015}
Schweinberger, M. and M.~S. Handcock (2015).
\newblock {Local dependence in random graph models: Characterization,
  properties and statistical inference}.
\newblock {\em Journal of the Royal Statistical Society: Series B (Statistical
  Methodology)\/}~{\em 77\/}(3), 647--676.

\bibitem[\protect\citeauthoryear{Snijders}{Snijders}{2002}]{Snijders2002}
Snijders, T. A.~B. (2002).
\newblock {Markov Chain Monte Carlo estimation of exponential random graph
  models}.
\newblock {\em Journal of Social Structure\/}.

\bibitem[\protect\citeauthoryear{Snijders, Pattison, Robins, and
  Handcock}{Snijders et~al.}{2006}]{snijders2006}
Snijders, T. A.~B., P.~E. Pattison, G.~L. Robins, and M.~S. Handcock (2006).
\newblock {New specifications for exponential random graph models}.
\newblock {\em Sociological Methodology\/}~{\em 36\/}(1), 99--153.

\bibitem[\protect\citeauthoryear{Stadtfeld, Tak{\'{a}}cs, and
  V{\"{o}}r{\"{o}}s}{Stadtfeld et~al.}{2020}]{Stadtfeld2020}
Stadtfeld, C., K.~Tak{\'{a}}cs, and A.~V{\"{o}}r{\"{o}}s (2020).
\newblock {The emergence and stability of groups in social networks}.
\newblock {\em Social Networks\/}~{\em 60}, 129--145.

\bibitem[\protect\citeauthoryear{Strauss and Ikeda}{Strauss and
  Ikeda}{1990}]{strauss1990pseudolikelihood}
Strauss, D. and M.~Ikeda (1990).
\newblock {Pseudolikelihood estimation for social networks}.
\newblock {\em Journal of the American Statistical Association\/}~{\em
  85\/}(409), 204--212.

\bibitem[\protect\citeauthoryear{Thurner, Schmid, Cranmer, and
  Kauermann}{Thurner et~al.}{2019}]{thurner2019network}
Thurner, P.~W., C.~S. Schmid, S.~J. Cranmer, and G.~Kauermann (2019).
\newblock Network interdependencies and the evolution of the international arms
  trade.
\newblock {\em Journal of Conflict Resolution\/}~{\em 63\/}(7), 1736--1764.

\bibitem[\protect\citeauthoryear{Wasserman and Faust}{Wasserman and
  Faust}{1994}]{Wasserman1994}
Wasserman, S. and K.~Faust (1994).
\newblock {\em {Social network analysis : Methods and applications}}.
\newblock Cambridge: Cambridge University Press.

\bibitem[\protect\citeauthoryear{Wasserman and Pattison}{Wasserman and
  Pattison}{1996}]{wasserman1996}
Wasserman, S. and P.~Pattison (1996).
\newblock {Logit models and logistic regressions for social networks: I. An
  introduction to Markov graphs andp}.
\newblock {\em Psychometrika\/}~{\em 61\/}(3), 401--425.

\end{thebibliography}


\begin{thebibliography}{}

\bibitem[\protect\citeauthoryear{Anders, Fariss, and Markowitz}{Anders
  et~al.}{2020}]{anders2020}
Anders, T., C.~J. Fariss, and J.~N. Markowitz (2020).
\newblock Bread before guns or butter: introducing surplus domestic product
  (sdp).
\newblock {\em International Studies Quarterly\/}~{\em 64\/}(2), 392--405.

\bibitem[\protect\citeauthoryear{Barndorff-Nielsen}{Barndorff-Nielsen}{1978}]{Barndorff-Nielsen1978}
Barndorff-Nielsen, O. (1978).
\newblock {\em {Information and exponential families in statistical theory}}.
\newblock New York: Wiley.

\bibitem[\protect\citeauthoryear{Gelman and Meng}{Gelman and
  Meng}{1998}]{Gelman1998}
Gelman, A. and X.~L. Meng (1998).
\newblock {Simulating normalizing constants: From importance sampling to bridge
  sampling to path sampling}.
\newblock {\em Statistical Science\/}~{\em 13\/}(2), 163--185.

\bibitem[\protect\citeauthoryear{Hage and Harary}{Hage and
  Harary}{1984}]{Hage_Harary_1984}
Hage, P. and F.~Harary (1984).
\newblock {\em Structural Models in Anthropology}.
\newblock Cambridge: Cambridge University Press.

\bibitem[\protect\citeauthoryear{Hummel, Hunter, and Handcock}{Hummel
  et~al.}{2012}]{Hummel2012}
Hummel, R.~M., D.~R. Hunter, and M.~S. Handcock (2012).
\newblock {Improving simulation-based algorithms for fitting ERGMs}.
\newblock {\em Journal of Computational and Graphical Statistics\/}~{\em
  21\/}(4), 920--939.

\bibitem[\protect\citeauthoryear{Hunter and Handcock}{Hunter and
  Handcock}{2006}]{hunter2006}
Hunter, D.~R. and M.~S. Handcock (2006).
\newblock {Inference in curved exponential family models for networks}.
\newblock {\em Journal of Computational and Graphical Statistics\/}~{\em
  15\/}(3), 565--583.

\bibitem[\protect\citeauthoryear{Kinne}{Kinne}{2020}]{Kinne_2020}
Kinne, B.~J. (2020).
\newblock The defense cooperation agreement dataset (dcad).
\newblock {\em Journal of Conflict Resolution\/}~{\em 64\/}(4), 729–755.

\bibitem[\protect\citeauthoryear{Krivitsky, Kuvelkar, and Hunter}{Krivitsky
  et~al.}{2022}]{Krivitsky}
Krivitsky, P.~N., A.~R. Kuvelkar, and D.~R. Hunter (2022).
\newblock {Likelihood-based Inference for Exponential-Family Random Graph
  Models via Linear Programming}.
\newblock {\em ArXiv\/}.
\newblock 2202.03572.

\bibitem[\protect\citeauthoryear{Lai and Reiter}{Lai and
  Reiter}{2000}]{Lai_Reiter_2000}
Lai, B. and D.~Reiter (2000).
\newblock Democracy, political similarity, and international alliances,
  1816-1992.
\newblock {\em Journal of Conflict Resolution\/}~{\em 44\/}(2), 203–227.

\bibitem[\protect\citeauthoryear{Lake}{Lake}{2009}]{Lake_2009}
Lake, D.~A. (2009).
\newblock {\em Hierarchy in International Relations}.
\newblock Ithaca: Cornell University Press.

\bibitem[\protect\citeauthoryear{Marshall, Gurr, and Jaggers}{Marshall
  et~al.}{2018}]{marshall2018polity}
Marshall, M.~G., T.~R. Gurr, and K.~Jaggers (2018).
\newblock Polity iv project: Political regime characteristics and transitions,
  1800--2017.
\newblock {\em Center for systemic peace\/}.

\bibitem[\protect\citeauthoryear{Miller}{Miller}{2021}]{peacesciencer}
Miller, S.~V. (2021).
\newblock \{peacesciencer\}: An r package for quantitative peace science
  research.
\newblock {\em Conflict Management and Peace Science\/}~{\em (OnlineFirst)}.

\bibitem[\protect\citeauthoryear{Palmer, McManus, D'Orazio, Kenwick, Karstens,
  Bloch, Dietrich, Kahn, Ritter, and Soules}{Palmer et~al.}{2021}]{Palmer2021}
Palmer, G., R.~W. McManus, V.~D'Orazio, M.~R. Kenwick, M.~Karstens, C.~Bloch,
  N.~Dietrich, K.~Kahn, K.~Ritter, and M.~J. Soules (2021).
\newblock The mid5 dataset, 2011–2014: Procedures, coding rules, and
  description.
\newblock {\em Conflict Management and Peace Science\/}~{\em 39\/}(4),
  470--482.

\bibitem[\protect\citeauthoryear{Read}{Read}{1954}]{read1954cultures}
Read, K.~E. (1954).
\newblock Cultures of the central highlands, new guinea.
\newblock {\em Southwestern Journal of Anthropology\/}~{\em 10\/}(1), 1--43.

\bibitem[\protect\citeauthoryear{Schoch}{Schoch}{2020}]{signnet}
Schoch, D. (2020).
\newblock {\em signnet: An R package to analyze signed networks}.
\newblock https://github.com/schochastics/signnet.

\bibitem[\protect\citeauthoryear{Schvitz, Girardin, R{\"u}egger, Weidmann,
  Cederman, and Gleditsch}{Schvitz et~al.}{2022}]{schvitz2022mapping}
Schvitz, G., L.~Girardin, S.~R{\"u}egger, N.~B. Weidmann, L.-E. Cederman, and
  K.~S. Gleditsch (2022).
\newblock Mapping the international system, 1886-2019: The cshapes 2.0 dataset.
\newblock {\em Journal of Conflict Resolution\/}~{\em 66\/}(1), 144--161.

\bibitem[\protect\citeauthoryear{Singer, Bremer, and Stuckey}{Singer
  et~al.}{1972}]{singer1972capability}
Singer, J.~D., S.~Bremer, and J.~Stuckey (1972).
\newblock Capability distribution, uncertainty, and major power war, 1820-1965.
\newblock In B.~Russett (Ed.), {\em Peace, war, and numbers}, pp.\  19--48.
  Beverly Hills: Sage.

\bibitem[\protect\citeauthoryear{Warren}{Warren}{2016}]{Warren_2016}
Warren, T.~C. (2016).
\newblock Modeling the coevolution of international and domestic institutions.
\newblock {\em Journal of Peace Research\/}~{\em 53\/}(3), 424–441.

\end{thebibliography}

\end{document}

\def\spacingset#1{\renewcommand{\baselinestretch}%
{#1}\small\normalsize} \spacingset{1}

%%%%%%%%%%%%%%%%%%%%%%%%%%%%%%%%%%%%%%%%%%%%%%%%%%%%%%%%%%%%%%%%%%%%%%%%%%%%%%

\if1\blind
{
  \title{\bf Supplementary Material $-$ \\  Exponential Random Graph Models for Dynamic Signed Networks: An Application to International Relations}
  \author{Cornelius Fritz\thanks{
    The authors gratefully acknowledge support from the German Federal Ministry of Education and Research (BMBF) under Grant No. 01IS18036A and the German Research Foundation (DFG) for the project TH 697/11-1: Arms Races in the Interwar Period 1919-1939. Global Structures of Weapons Transfers and Destabilization. }\hspace{.2cm}\\
    Department of Statistics, LMU Munich\\
    Marius Mehrl \\
    Paul W. Thurner \\
    Geschwister Scholl Institute of Political Science \\
    Göran Kauermann\\
    Department of Statistics, LMU Munich
    }
  \maketitle
} \fi

\if0\blind
{
  \bigskip
  \bigskip
  \bigskip
  \begin{center}
    {\LARGE\bf  Supplementary Material:  Exponential Random Graph Models for Dynamic Signed Networks: An Application to International Relations}
\end{center}
  \medskip
} \fi

%\spacingset{1.9} % DON'T change the spacing!

\setcounter{equation}{0}
\section{Technical Details}
\subsection{Partial Stepping Algorithm}
\label{sec:stepping}

Following standard theory of exponential families, $\theta$ maximizing the approximate likelihood detailed in (11) of the main article only exists if the observed sufficient statistics  $\sum_{t = 1}^T \mathbf{s}(\mathbf{y}_t, \mathbf{y}_{t-1})$ are inside the convex hull spanned by the sampled sufficient statistics \newline $\left(\sum_{t = 1}^T \mathbf{s}(\mathbf{y}_t^{(m)}, \mathbf{y}_{t-1}), ..., \sum_{t = 1}^T \mathbf{s}(\mathbf{y}_t^{(M)}, \mathbf{y}_{t-1})\right)$ \citepsupp{Barndorff-Nielsen1978}. Since this condition does not hold for arbitrary values of $\theta_0$, we adapt the partial stepping algorithm introduced by \citetsupp{Hummel2012} to find an adequate $\theta_0$.    

In the $k$th step of this iterative procedure, we substitute $\sum_{t = 1}^T \mathbf{s}(\mathbf{y}_t, \mathbf{y}_{t-1})$ in (11) of the main article by 
\begin{align}
\label{eq:xi}
 \xi^{(k)} = \gamma^{(k)} \sum_{t = 1}^T \mathbf{s}(\mathbf{y}_t, \mathbf{y}_{t-1})+ \left(1-\gamma^{(k)}\right)\widehat{\mathbf{m}}^{(k)},   
\end{align}
where $\gamma^{(k)} \in (0,1]$ and $\widehat{\mathbf{m}}^{(k)} = \frac{1}{M} \sum_{m = 1}^M \sum_{t = 1}^T \mathbf{s}(\mathbf{y}_t^{(m)}, \mathbf{y}_{t-1})$ is the estimated mean of the sufficient statistics of networks sampled under $\theta^{(k)}$. We select the largest possible value of $\gamma^{(k)}$ in \eqref{eq:xi} such that even the point marginally closer to $\sum_{t = 1}^T \mathbf{s}(\mathbf{y}_t, \mathbf{y}_{t-1})$, defined by $1.05\gamma^{(k)} \sum_{t = 1}^T \mathbf{s}(\mathbf{y}_t, \mathbf{y}_{t-1}) + \left(1-1.05\gamma^{(k)}\right) \mathbf{m}^{(k)}$, is inside the convex hull spanned by the sampled statistics. One can test whether a point $\in \mathbb{R}^p$ lies in this convex hull via a linear programming problem (details can be found in \citealpsupp{Hummel2012} and \citealpsupp{Krivitsky}). 

To update $\theta^{(k)}$ to $\theta^{(k+1)}$ for a given $\gamma^{(k)}$, we thus optimize
\begin{equation}
\begin{aligned}
 &  \left(\theta^{(k+1)} - \theta^{(k)}\right)^\top  \xi^{(k)} -\log\left(\frac{1}{M} \sum_{m = 1}^M\exp\left\{(\theta - \theta_0)^\top \left(\sum_{t = 1}^T \mathbf{s}(\mathbf{y}_t^{(m)}, \mathbf{y}_{t-1})\right)\right\}\right),    \label{eq:lhr_approx_2}
\end{aligned}
\end{equation}
with a Newton-Raphson algorithm. To ease this step, we assume that $\sum_{t = 1}^T\mathbf{s}(\mathbf{Y}_t, \mathbf{y}_{t-1})$ follows a $p$-variate Gauss distribution with mean $\mathbf{m}^{(k)}$ and covariance matrix $\mathbf{\Sigma}^{(k)}$, which is the covariance matrix of the sufficient statistics under $\theta^{(k)}$. Both terms can be estimated with samples  $\mathbf{Y}^{(1)}, ..., \mathbf{Y}^{(M)}$. Thereby we can state the optimal value of \eqref{eq:lhr_approx_2} in closed form: 
\begin{align*}
   \theta^{(k+1)} = \theta^{(k)} + \left(\widehat{\mathbf{\Sigma}}^{(k)}\right)^{-1}\left(\xi^{(k)} - \widehat{\mathbf{m}}^{(k)}\right).
\end{align*}
%This adjustment is equivalent to the Robins-Monroe algorithm as proposed by \citet{Snijders2002} if we update the variance of the simulated statistics in each iteration and under constant step length. 
The algorithm terminates when we estimate $\gamma^{(k)} = 1$ two iterations in a row, we then continue the procedure with $\xi^{(k)} = \sum_{t = 1}^T \mathbf{s}(\mathbf{y}_t, \mathbf{y}_{t-1})$ until the estimates stabilize.  

\label{sec:mod_sel}
%\subsection{Evaluation of the AIC}
\subsection{Evaluation of the AIC}

To decide between alternative specifications of the sufficient statistics, a common method is to select the model with the lowest AIC value.  The AIC is defined as 
\begin{align}
\label{eq:aic}
    \text{AIC}(M) = 2p - 2 \ell(\hat{\theta};\mathbf{y}),
\end{align}
where $M$ is a SERGM for temporal networks with a particular specification of the sufficient statistics and estimated parameters $\hat{\theta}$ and $\ell(\hat{\theta};\mathbf{y}) = \log\left(\prod_{t = 1}^T \mathbb{P}_\theta (\mathbf{Y}_t = \mathbf{y}_t |\mathbf{Y}_{t-1} = \mathbf{y}_{t-1})\right)$ is the log likelihood. To evaluate \eqref{eq:aic}, we have to calculate the value of the intractable logarithmic likelihood at $\hat{\theta}$, which we can restate by 
\begin{align}
    \label{eq:lh_approx_5}
    \ell(\hat{\theta};\mathbf{y}) = r(\hat{\theta}, \hat{\theta}_{\text{Ind}};\mathbf{y}) +\ell( \hat{\theta}_{\text{Ind}};\mathbf{y}),
\end{align}
where $\hat{\theta}_{\text{Ind}} \in \mathbb{R}^p$ is the estimate of the sub-model including only the subset from the sufficient statistics that abide the conditional dependence assumption (the coefficients of all other (endogenous) statistics are fixed at 0). Due to this characteristic, $\ell( \hat{\theta}_{\text{Ind}};\mathbf{y})$ is equivalent to the log likelihood in a multinomial regression and can be computed in closed form. To evaluate $r(\hat{\theta}, \hat{\theta}_{\text{Ind}};\mathbf{y})$, we follow \citetsupp{hunter2006} and apply path sampling \citepsupp{Gelman1998} to approximate 
\begin{align*}
\log\left(\mathbb{E}_{\theta_0}\left(\exp\left\{(\theta - \theta_0)^\top \left(\sum_{t = 1}^T \mathbf{s}(\mathbf{Y}_t, \mathbf{y}_{t-1})\right)\right\}\right)\right) = \log\left(\frac{\kappa(\theta)}{\kappa(\theta_0)}\right)    
\end{align*}
from (11) in the main article in a more precise manner. If we specify a smooth mapping $\theta\text{ : }[0,1]\rightarrow\mathbb{R}^p$ with $\theta(0) = \hat{\theta}_{\text{Ind}}$ and $\theta(1) = \hat{\theta}$ and let $0= u_0<u_1 < ...< u_J = 1$ for $J\in \mathbb{N}$ be a fixed grid of {so-called} bridges and its finite support, the following approximation holds:  
\begin{align}
    \label{eq:bridge_sample}
    \log\left(\frac{\kappa(\theta)}{\kappa(\theta_0)}\right) &= \sum_{j = 1}^J \frac{1}{u_j - u_{j-1}} \mathbb{E}_{\theta(u_j)}\left( \left(\frac{\text{d}}{\text{d} u} \boldsymbol{\theta}(u)\bigg|_{u = u_j}\right)^\top  \sum_{t = 1}^T \mathbf{s}\left(\mathbf{Y}_t, \mathbf{y}_{t-1} \right) \right)\nonumber\\ &\approx  \frac{1}{M} \sum_{j = 1}^J \sum_{m = 1}^M \frac{1}{u_j - u_{j-1}} \left(\left(\frac{\text{d}}{\text{d} u} \boldsymbol{\theta}(u)\bigg|_{u = u_j}\right)^\top \sum_{t = 1}^T \mathbf{s}\left(\mathbf{y}^{(j,m)}_t, \mathbf{y}_{t-1} \right)\right),
\end{align}
where $\mathbf{y}^{(j,1)}_t, ..., \mathbf{y}^{(j,M)}_t$ are networks sampled conditional on $\mathbf{y}_{t-1}$ under $\theta(u_j)~\forall~ j = 1, ..., J$. For our implementation, we set $\theta(u) = \hat{\theta}_{\text{Ind}} + u(\hat{\theta}-\hat{\theta}_{\text{Ind}})$, corresponding to a linear path from $\hat{\theta}_{\text{Ind}}$ to $\hat{\theta}$ and $ \frac{\text{d}}{\text{d} u} \theta(u)= \hat{\theta}-\hat{\theta}_{\text{Ind}}$. Plugging \eqref{eq:bridge_sample} into the first row of (11) of the main article  permits the computation of \eqref{eq:aic}. For a more technical derivation of \eqref{eq:bridge_sample}, we refer to \citealpsupp{hunter2006} or \citealpsupp{Gelman1998}.

\section{Details on the Application to International Cooperation and Conflict}

\subsection{Data Visualization and Covariate Details}

\begin{figure}[t!]
		\centering
		\includegraphics[width=0.8\linewidth,page =1 ]{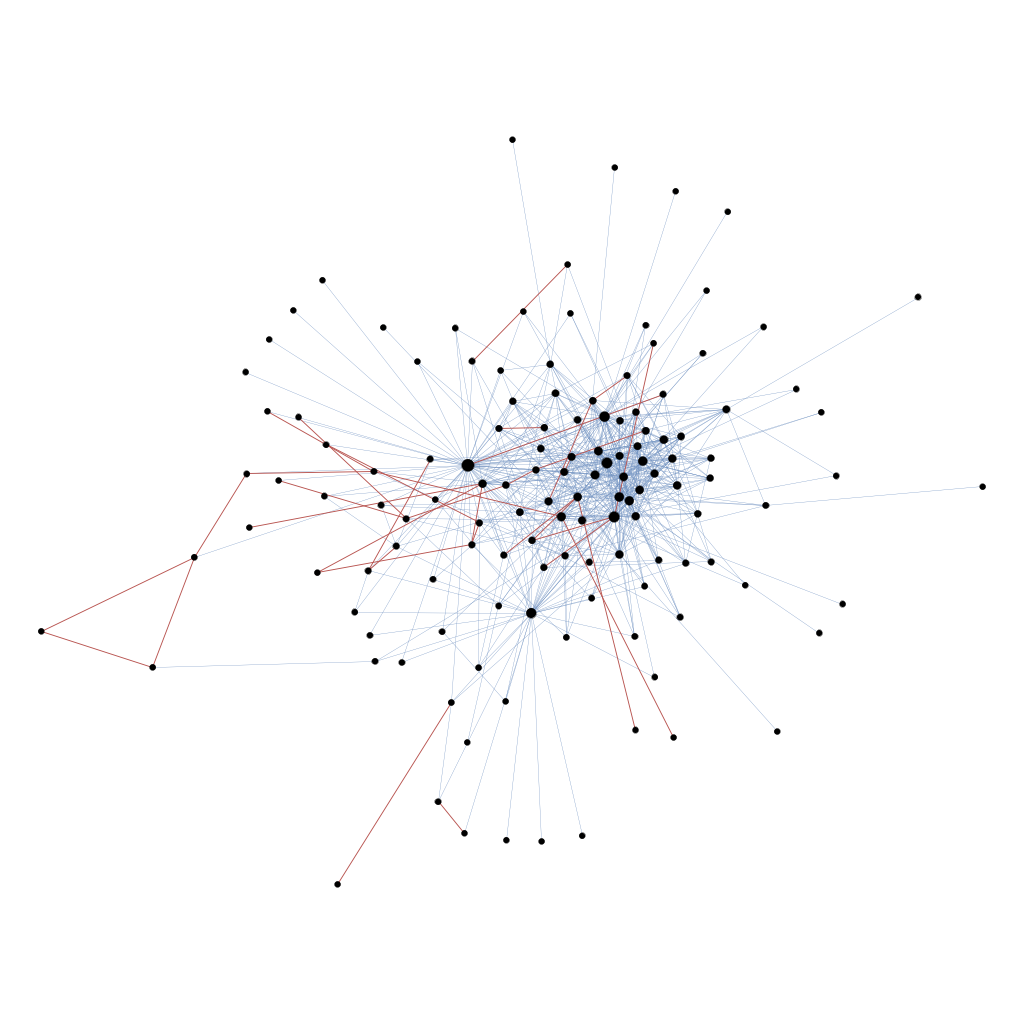}
		\caption{Network of MIDs (red) and DCAs (blue) in 2005. The size of each node relates to the degree (positive plus negative) of the respective country. 
		}
				\label{fig:network_states}
\end{figure}

Here, we visualize the network and offer additional details regarding the data sources for the application of the SERGM to interstate relations presented in Section 4. As discussed in this section, we source data from \citetsupp{Kinne_2020} and \citetsupp{Palmer2021} to construct a network, spanning the years 2000-2010, where positive ties represent Defense Cooperation Agreements (DCAs) and negative ties Militarized Interstate Disputes (MIDs) between states. A snapshot of the resulting network, as observed in 2005, is presented in Figure \ref{fig:network_states}.    

For this application, we also use additional data to construct our exogenous covariates. The information underlying these variables, as well as the MID data, are sourced from the $\mathtt{peacesciencer}$ package \citepsupp{peacesciencer}, but the original data sources are as follows: We measure countries' absolute political difference (Abs. Polity Diff.) using their polity scores \citepsupp{marshall2018polity}, their relative military power by taking the ratio of their Composite Indicators of National Capabilities\footnote{We use the higher CINC value in the ratio's numerator.} \citepsupp[CINC Ratio; ][]{singer1972capability}, their difference in wealth via their absolute GDP difference \citepsupp[Abs. GDP Diff; ][]{anders2020} and obtain their geographical distance from \citetsupp{schvitz2022mapping}, log-transforming it before inclusion (Abs. Distance). For each covariate, we separately estimate effects on the propensity of a positive and negative edge. The only exception to this rule is the effect of the absolute distance, which is assumed to be equal for both types of edges.% We also incorporate the state of the ne %We derive pairwise scalar information by following the suggestion of Section 2.2 of the main article and using the absolute 

We now shortly discuss the estimation results for these covariates, as reported for Model 1 in Table 1 of the main article. These estimates are ceteris paribus, i.e. when accounting for network dependencies via the endogenous terms. Regarding cooperation, countries are found to be more likely to formally work together via defense cooperation agreements if they are politically more similar, more comparable in their wealth, but also differ more in their material military capabilities. In particular the first result is in line with previous research showing that similar regimes are more likely to ally \citepsupp{Lai_Reiter_2000,Warren_2016} while the second indicates that for DCAs, which regulate activities such as the joint research and development of military technology, countries' economic match also plays a role. That countries are more likely to cooperate as their CINC ratio increases indicates, instead, that DCAs also follow a hierarchical structure where powerful states enter agreements with less powerful ones \citepsupp{Lake_2009}. In contrast, we see that states are more likely to fight if their CINCs, and hence military capabilities, are more similar whereas their differences in terms of regime type and wealth are not found to play a role. Finally, countries' absolute distance exhibits a positive coefficient, indicating that, surprisingly, they are more likely to interact the farther they are away from each other.

\subsection{MCMC Diagnostics}

\begin{figure}[t!]\centering
	 \centering
    \includegraphics[width=\linewidth]{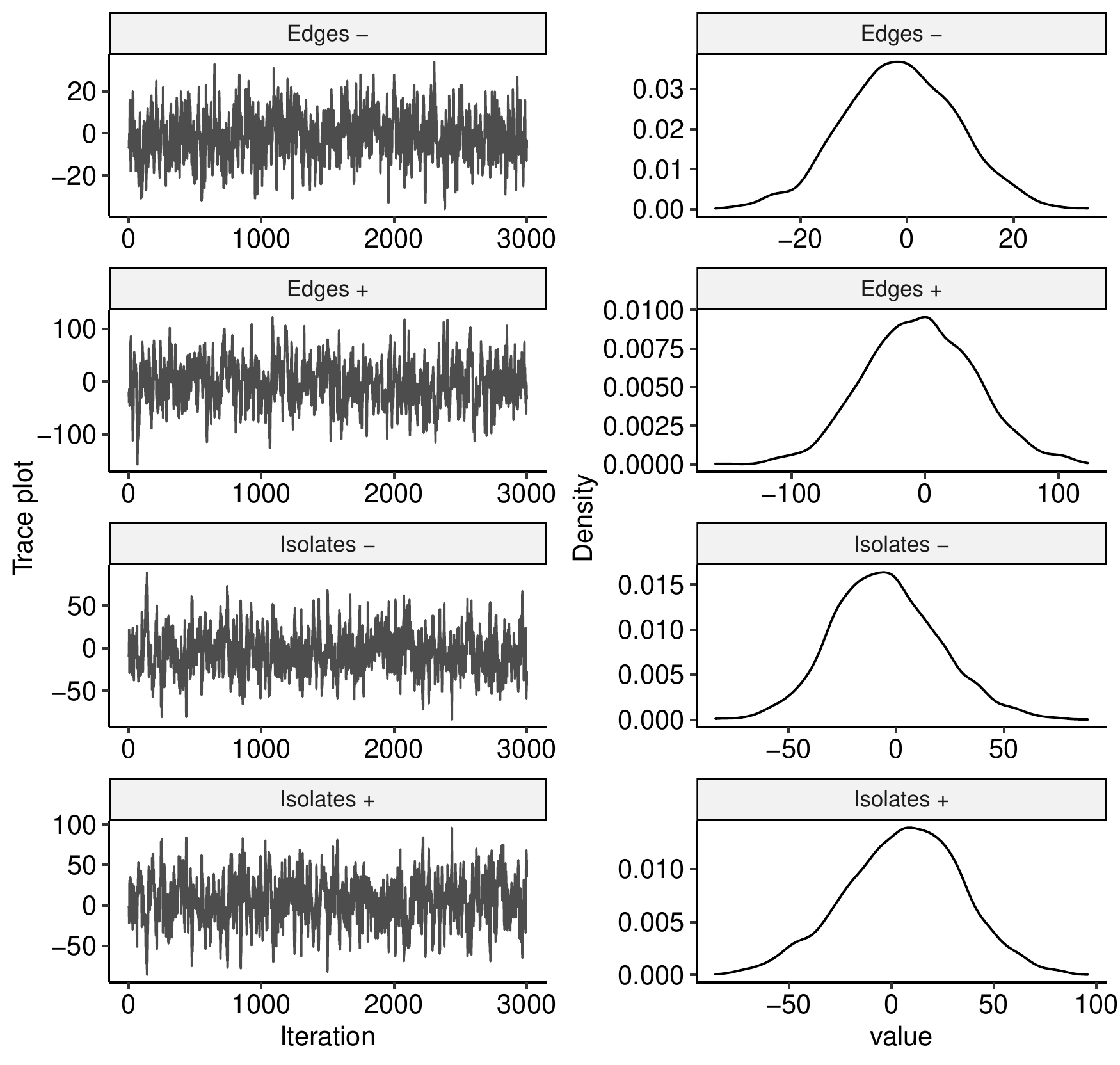}
	\caption{MCMC diagnostics of Model 1.}
	\label{fig:trace_cow1}
\end{figure}

In figures \ref{fig:trace_cow1}--\ref{fig:trace_cow4}, we present some diagnostic plots of the MCMC chain used in the final iteration of Model 1 in the application of Section 4 of the main article. We average the Markov chain of each sufficient statistic around its observed value for better readability. Overall, one can observe that the model's estimates converged, are not degenerate, and are equal to the maximum likelihood estimates since the Markov chain oscillates around 0.

\begin{figure}[t!]\centering
	 \centering
    \includegraphics[width=\linewidth]{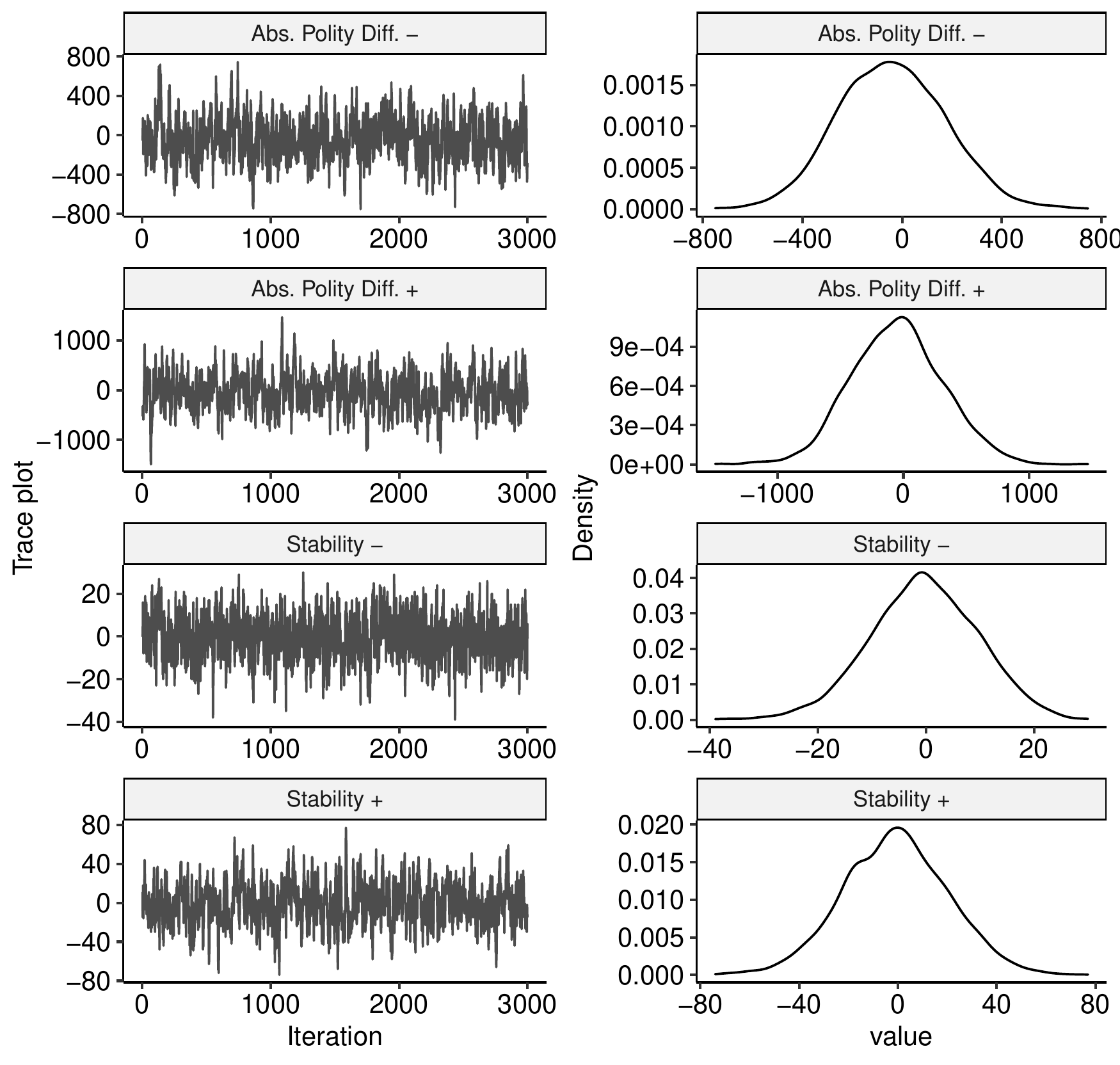}
	\caption{MCMC diagnostics of Model 1.}
	\label{fig:trace_cow2}
\end{figure}

\begin{figure}[t!]\centering
	 \centering
    \includegraphics[width=\linewidth]{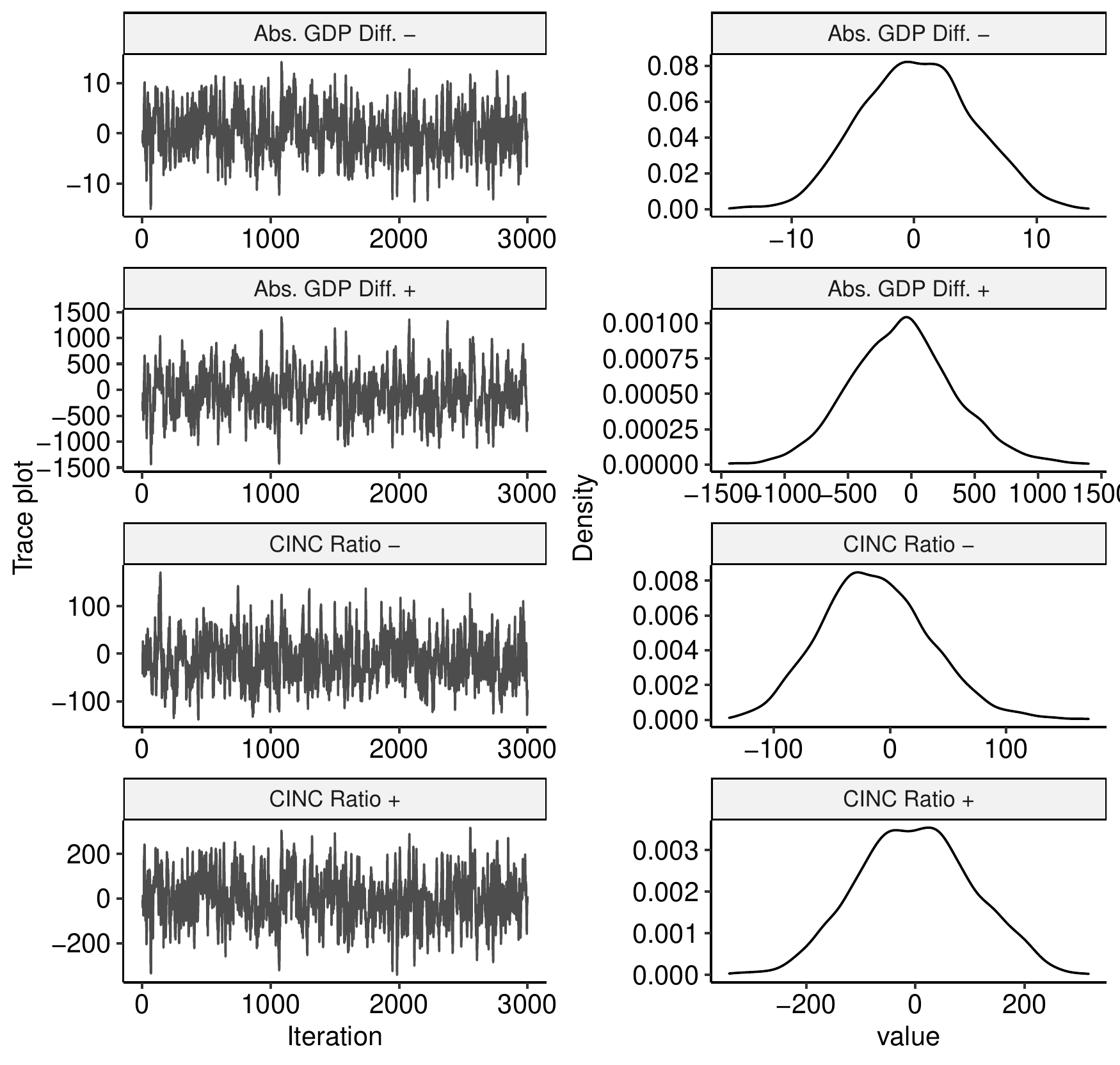}
	\caption{MCMC diagnostics of Model 1.}
	\label{fig:trace_cow3}
\end{figure}

\begin{figure}[t!]\centering
	 \centering
    \includegraphics[width=\linewidth]{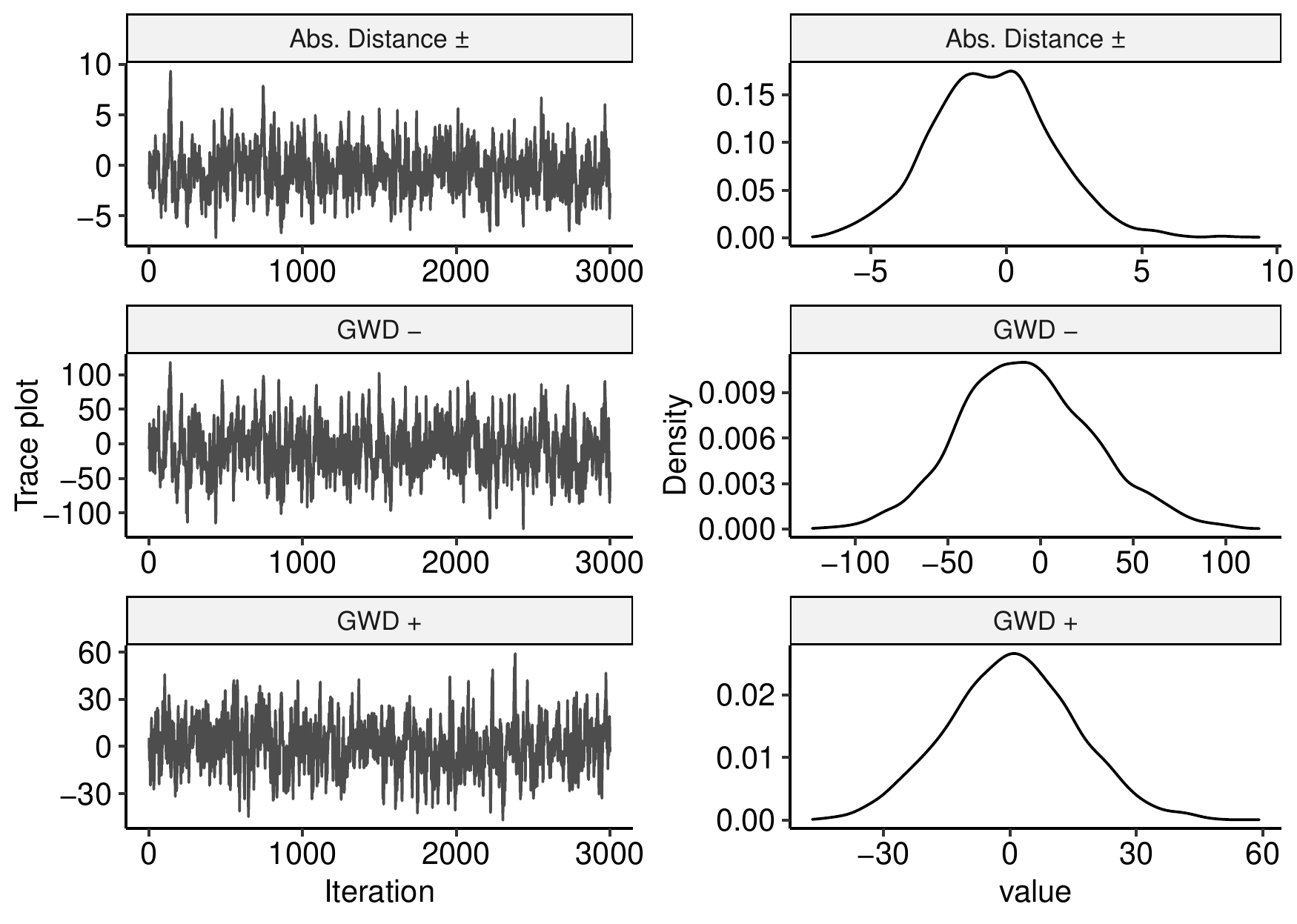}
	\caption{MCMC diagnostics of Model 1.}
	\label{fig:trace_cow4}
\end{figure}
\begin{figure}[t!]\centering
	 \centering
    \includegraphics[width=\linewidth]{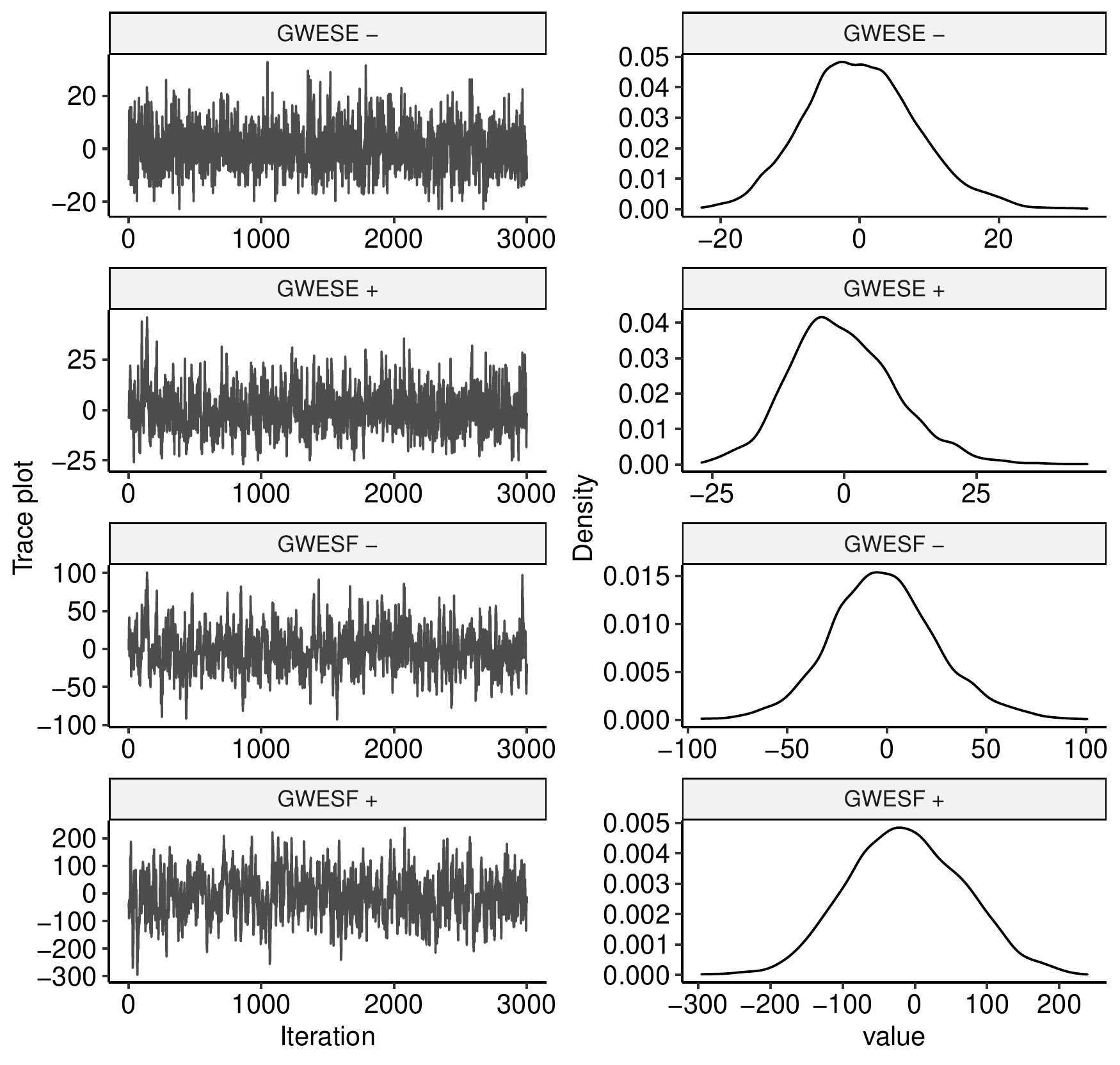}
	\caption{MCMC diagnostics of Model 1.}
	\label{fig:trace_cow4}
\end{figure}
\section{Application to a static network: Enmity and Friendship among New Guinean Highland Tribes}
\subsection{Data Visualization and Sources}

\begin{figure}[t!]
		\centering
		\includegraphics[width=0.7\linewidth,page =1 ]{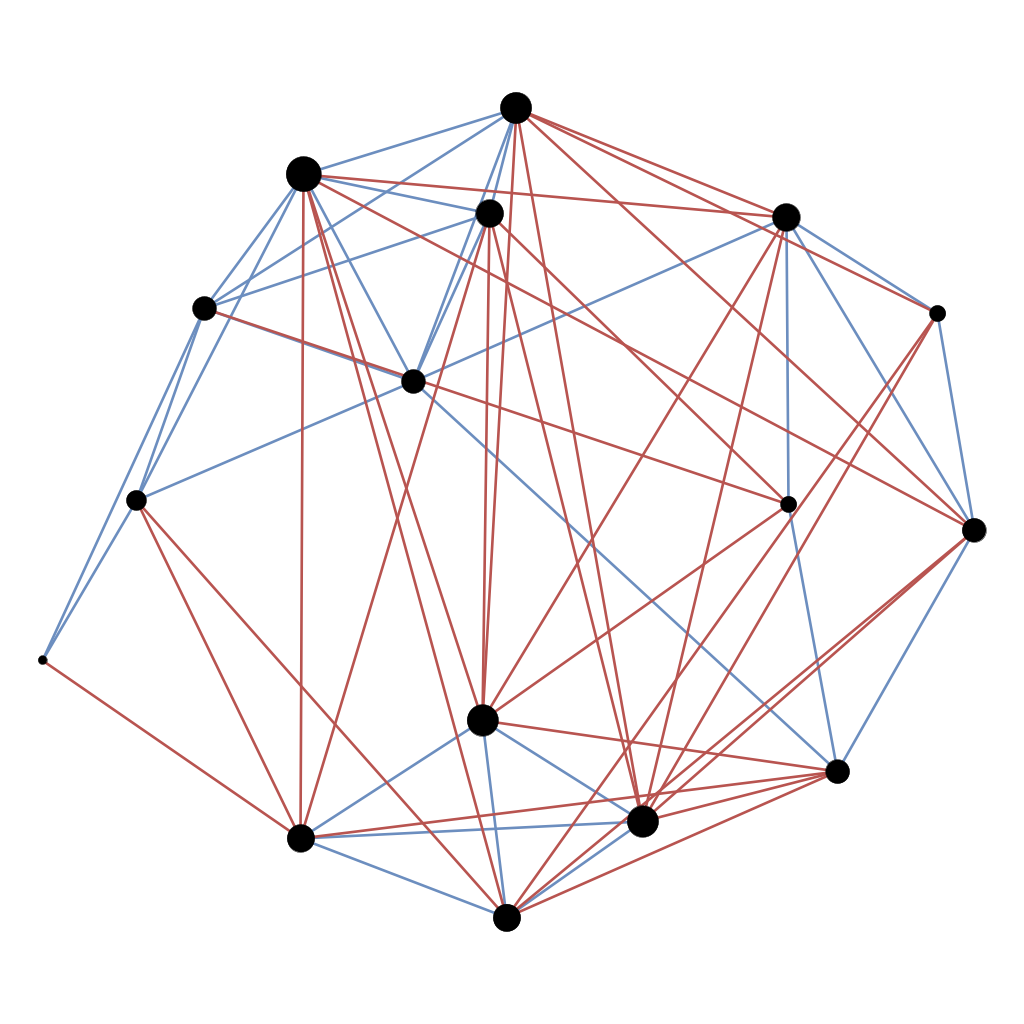}
		\caption{Network of enmity (red) and friendship (red) among New Guinean Highland tribes. The size of each node relates to the degree (positive plus negative) of the respective people.}
				\label{fig:network_tribes}
\end{figure}

Next to dynamic networks one can also apply the SERGM to static networks. We demonstrate this with network data on interactions between the New Guinean Highland Tribes originally collected by \citetsupp{read1954cultures} and presented in \citetsupp{Hage_Harary_1984}. We source these data from the $\mathtt{R}$ package $\mathtt{signnet}$ \citepsupp{signnet}. The network covers relations of enmity and friendship among sixteen subtribes of the Gahuku-Gama, based on the anthropological work of \citetsupp{read1954cultures}. \citetsupp{Hage_Harary_1984} introduce it as an example of a network which is not perfectly balanced due to the existence of triads with zero or two positive ties but note that 82\% of triads are balanced nonetheless. The full network is plotted in Figure \ref{fig:network_tribes}. We now apply the SERGM to this static network to test whether structural balance effects can be recovered from it. The SERGM we specify includes edge terms, $GWESE$, $GWESF$ as well as a degree statistic. To show the flexibility with which these statistics can be specified, we include the edge and $GWESF$ terms separately for positive and negative ties, but the $GWESE$ and $GWD$ statistics only for positive ties. For the sake of comparison, we also estimate a model that drops all endogenous network terms and hence includes only the two edge terms. Results of both models are presented in Table \ref{tab:highland}.

\subsection{Results and Model Assessment}

\begin{table}[!tbp]
\caption{Results of the models.\label{tbl:res_cow}} 
\begin{center}
\begin{tabular}{lllcll}
\hline
\multicolumn{1}{l}{\  }&\multicolumn{2}{c}{\  Dependence}&\multicolumn{1}{c}{\  }&\multicolumn{2}{c}{\  Independence}\tabularnewline
\cline{2-3} \cline{5-6}
\multicolumn{1}{l}{}&\multicolumn{1}{c}{Coef.}&\multicolumn{1}{c}{CI}&\multicolumn{1}{c}{}&\multicolumn{1}{c}{Coef.}&\multicolumn{1}{c}{CI}\tabularnewline
\hline
Edges +&-8.744&[-12.182,-5.306]&&-0.76&[-1.13,-0.39]\tabularnewline
Edges $-$&-1.647&[-2.766,-0.528]&&-0.76&[-1.13,-0.39]\tabularnewline
$GWESE^+$&0.45&[0.015,0.885]&&-&\tabularnewline
$GWESF^+$&0.068&[-0.228,0.364]&&-&\tabularnewline
$GWESF^-$&0.932&[0.616,1.248]&&-&\tabularnewline
$GWD^+$&5.492&[2.252,8.732]&&-&\tabularnewline
\hline

AIC&139.593&&&170.355&\tabularnewline
\hline
\end{tabular}     \label{tab:highland}
\end{center}
\end{table}

In Table \ref{tab:highland}, it is apparent that the fully specified Dependence model has a lower AIC than the Independence model which does not account for endogenous network terms, indicating that it is preferable in terms of performance. Table \ref{tab:highland} also offers some evidence that structural balance drives tie formation among the Gahuku-Gama: $GWESF$ has a positive effect on positive ties whose 95\%-Confidence Intervals clearly exclude zero  while its effect on negative ties is very close to zero. This implies that here, friends of friends are indeed more often friends but not less often enemies than one would expect by chance. $GWESE^+$ also exhibits a positive and statistically significant effect, suggesting that subtribes with a common enemy are more likely to share an alliance than in a random network of the same size. Finally, the effect of $GWD^+$ is also positive and statistically significant, meaning that a subtribe's probability of gaining a further positive tie increases with the number of such ties it already has.

\begin{figure}[t!]
		\centering
		\includegraphics[width=0.8\linewidth,page =1 ]{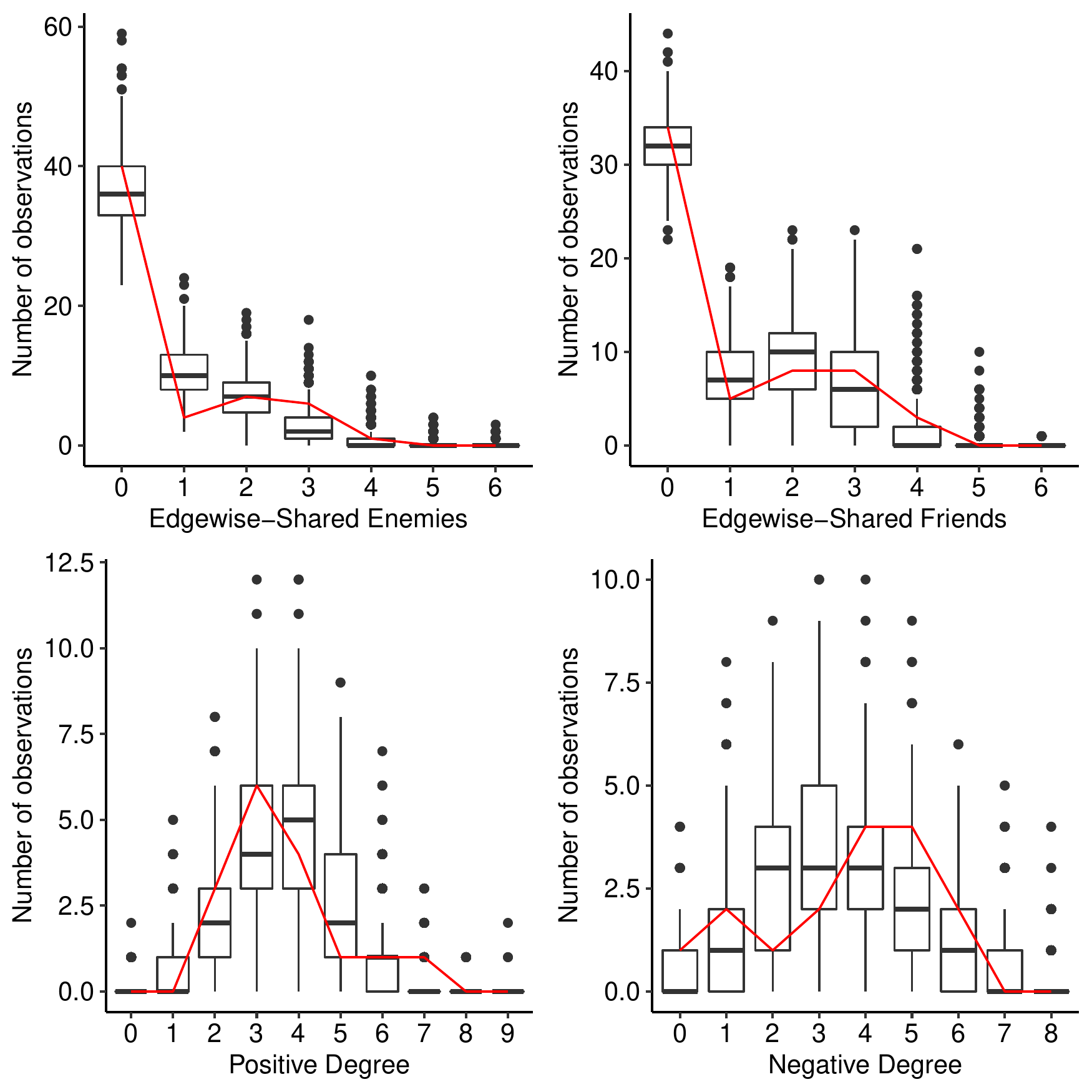}
		\caption{Model Assessment of the New Guinean Highland Tribes.
		}
				\label{fig:mod_assessment_tribes}
\end{figure}

Figure \ref{fig:mod_assessment_tribes} offers a visual assessment of the goodness-of-fit of the Dependence model. Here, we can see that while the observed network lines up quite well with the simulated ones in terms of Edgewise-Shared Friends, model fit is more problematic for Edgewise-Shared Enemies where the observed network is regularly outside the interquartile range of the simulated networks. Similarly, the model does not do a good job of capturing the observed network's negative degree distribution. Based on these plots, one may consider re-running the Dependence model while specifying $GWESE$ for both positive and negative ties and including $GWD^-$. Nonetheless this application demonstrates the possibility to use the SERGM for the analysis of static networks. 

\subsection{MCMC Diagnostics}
\begin{figure}[t!]\centering
	 \centering
    \includegraphics[width=\linewidth]{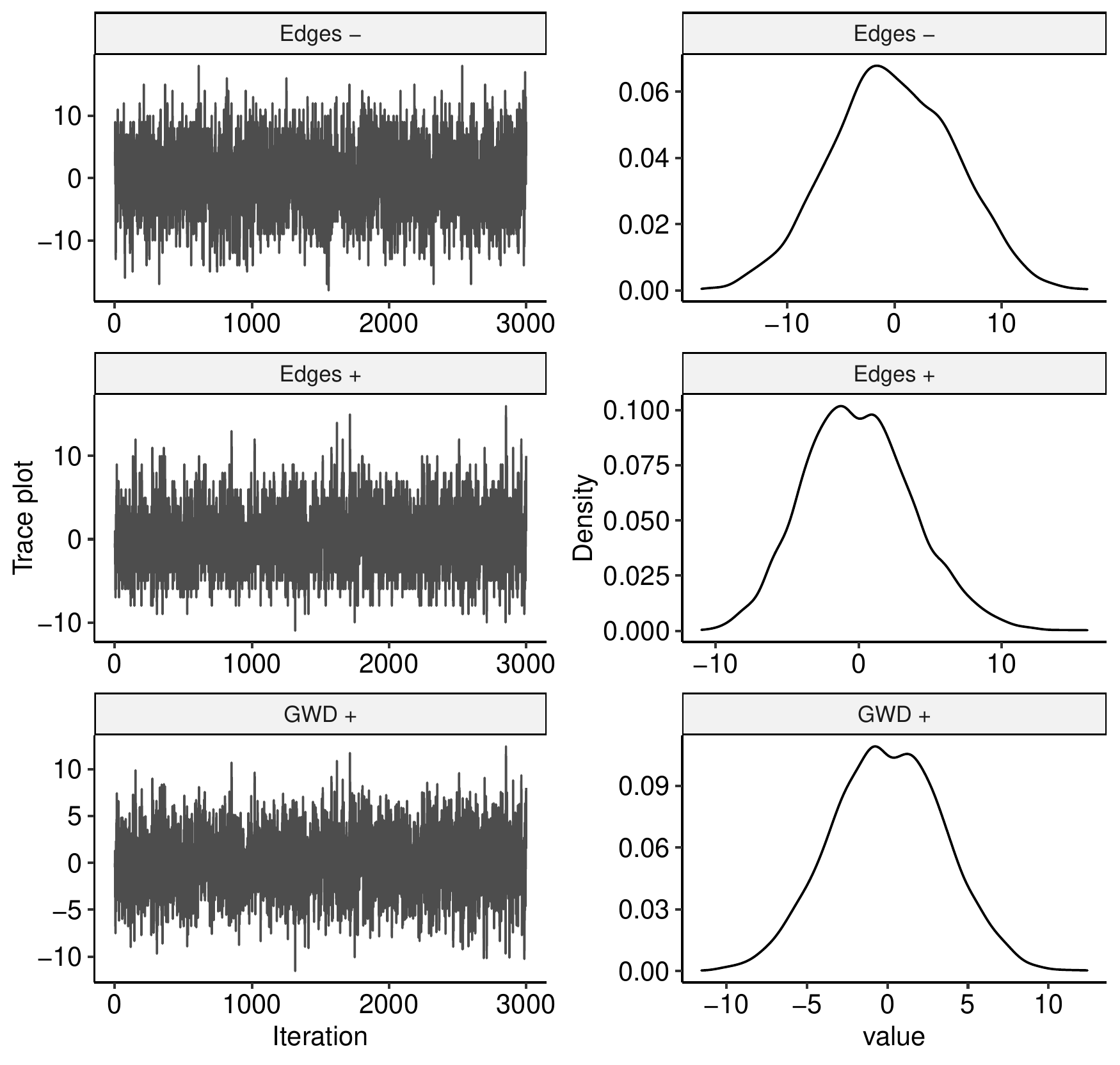}
	\caption{MCMC diagnostics of Dependence Model.}
	\label{fig:trace_tribe1}
\end{figure}

\begin{figure}[t!]\centering
	 \centering
    \includegraphics[width=\linewidth]{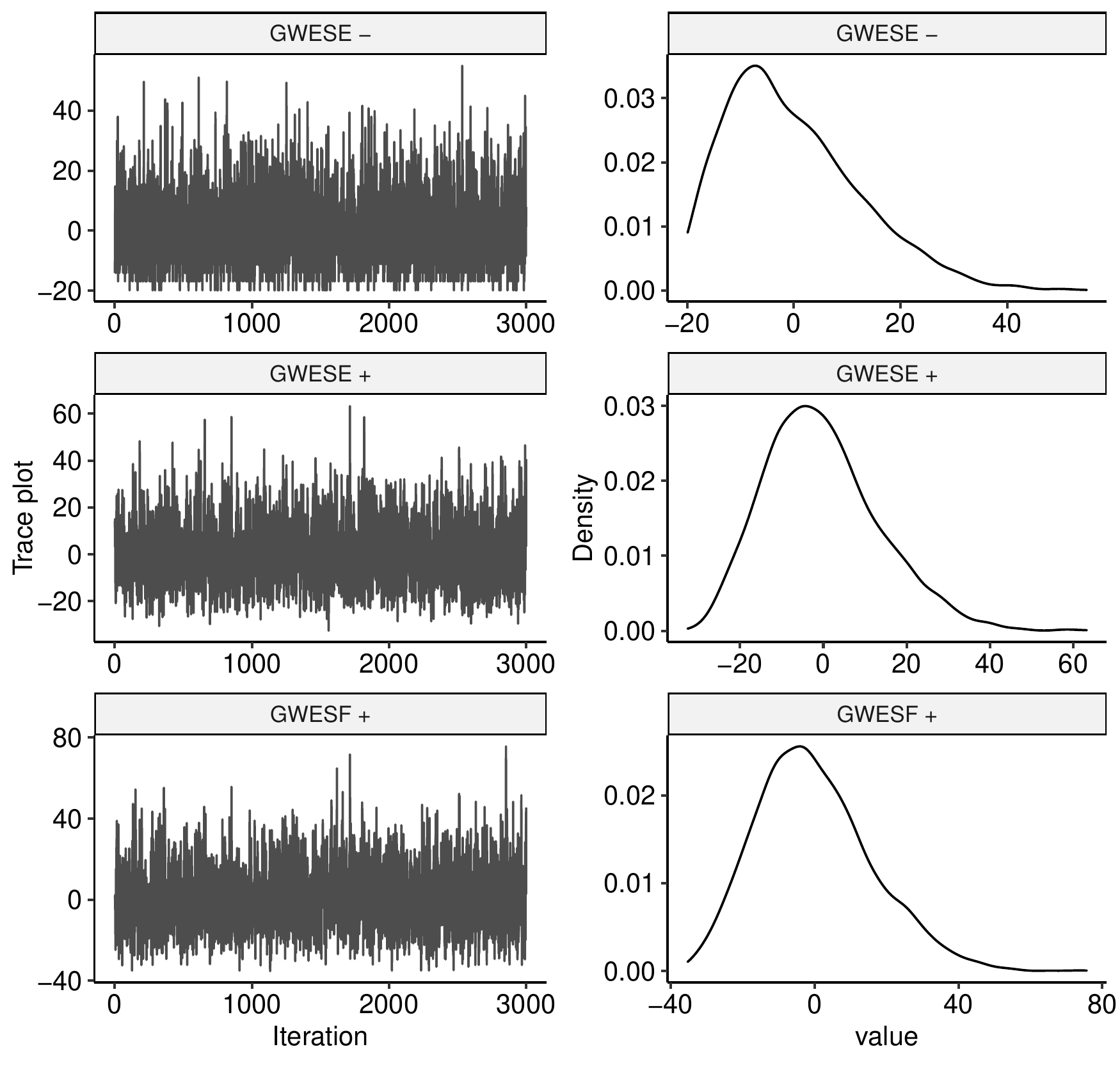}
	\caption{MCMC diagnostics of Dependence Model.}
	\label{fig:trace_tribe2}
\end{figure}

Finally, we also present the MCMC diagnostics for this additional application. Below are thus shown the MCMC trace plots for all its covariates (Figures \ref{fig:trace_tribe1}--\ref{fig:trace_tribe2}). Overall, these MCMC diagnostics indicate a good convergence of the model.

\section{Computational Settings}

We provide the tuning parameters of the MCMC algorithm and the models for both applications in Table \ref{tbl:setting}. We performed sensitivity checks to guarantee that the reported findings do not depend on the fixed parameters. In general, the values should depend on the density (the lower to higher the burn-In and MCMC interval), size (the higher to higher the burn-In and MCMC interval), and the strength of exogenous covariates (the higher to higher the burn-In and MCMC interval) of the network.  

\begin{table}[t!]
\center
\begin{tabular}{ll|ll}
                 &                         & Cooperation and Conflict &  Tribes \\\hline
                 & Fixed $\alpha$                         &      $\log(2)$                                  &       1.5                                                  \\
& Grid size for $\gamma$ &       3,000                                 &         2,000                                                \\
& Number of Bridges                         &      16                                  &       16                                                  \\\hline
  \multirow{4}{*}{\begin{sideways}Estimation\end{sideways}}  & Burn-In                                   &     10,000                                   &                   10,000                                      \\
& MCMC Interval                             &            1,000                             &     1,000                                                    \\
& M                                         &      2,000                                  &       1,000                                                  \\
& Start Empty                               &           False                             &       False                                                 \\\hline
   \multirow{4}{*}{\begin{sideways}Variance\end{sideways}}  & Burn-In                                   &                                  100,000      &                                  10,000                       \\
& MCMC Interval                             &       10,000                                 &             1,000                                            \\
& M                                         &       3,000                                 &          3,000                                               \\
& Start Empty                               &        True                                &        False                                                \\\hline
  \multirow{4}{*}{\begin{sideways}Bridge \end{sideways}}  & Burn-In                                   &                                    10,000    &                10,000                                         \\
& MCMC Interval                             &                            1,000            &                    2,000                                     \\
& M                                         &                3,000                        &       1,000                                                  \\
& Start Empty                               &         True                               &             True                                           
\end{tabular}
\caption{Setting of the parameters that need to be set during the fitting of the MCMC estimation procedure. One can define three different samplers, for the estimation of the parameters, for the quantification of their variance, and for the evaluation of the AIC.}
\label{tbl:setting}
\end{table}
%section{Tutorial for $\mathtt{ergm.sign}$ Package}
%\section{Implementation: The $\mathtt{ergm.sign}$ Package}

\bibliographystylesupp{chicago}
\bibliographysupp{references}